\documentclass[sigconf, screen, nonacm]{acmart}

\AtBeginDocument{%
  }

\setcopyright{none} 
\renewcommand\footnotetextcopyrightpermission[1]{}

\usepackage{amsmath, array, arydshln, subcaption, algorithm,algpseudocode}
\usepackage{physics}
\usepackage{subcaption}
\usepackage{tikz}
\usetikzlibrary{arrows.meta, positioning, shapes.geometric}

\usepackage{amsmath, array, arydshln, algorithm, algpseudocode, tikz,longtable}
\usetikzlibrary{arrows.meta, positioning, shapes.geometric}
\usepackage{physics}
\usepackage{subcaption}
\usepackage{enumitem}

\usepackage{xspace}
\newcommand{\framework}{STABSim\xspace}

\author{Sean Garner}
\affiliation{
  \institution{University of Washington, \\ Pacific Northwest National Lab}
    \country{USA}
}
\email{seangarn@uw.edu}
  
\author{Chenxu Liu}
\affiliation{
  \institution{Pacific Northwest National Lab}
\country{USA}
}
\email{chenxu.liu@pnnl.gov}

\author{Meng Wang}
\affiliation{
  \institution{Pacific Northwest National Lab,\\ University of British Columbia}
  \country{USA}
}
\email{meng.wang@pnnl.gov}

\author{Samuel Stein}
\affiliation{%
  \institution{Pacific Northwest National Lab}
  \country{USA}
}
\email{samuel.stein@pnnl.gov}

\author{Ang Li}
\affiliation{%
  \institution{Pacific Northwest National Lab, \\ University of Washington}
  \country{USA}
}
\email{ang.li@pnnl.gov}

\begin{document}

\title{STABSim: A Parallelized Clifford Simulator with Features Beyond Direct Simulation}

\begin{abstract}
The quantum stabilizer formalism became foundational for understanding error correction soon after the realization of the first useful quantum error correction codes. Stabilizers provide a way to describe sets of quantum states which are valid codewords within a quantum error correction (QEC) scheme. Existing stabilizer simulators are single threaded applications used to sample larger codes than is possible with other methods. However, there is an outstanding gap in the scaling and accuracy of current simulators for QEC as quantum computing exceeds hundreds of qubits, along with an under-utilization of the capabilities of highly-efficient stabilizer simulation across other quantum domains. In this work, we present the first GPU-accelerated tableau stabilizer simulator to scale better than CPU methods in QEC workloads, by trivializing Clifford gates and exploiting the large parallelism of dedicated GPUs with CUDA warp-level primitives to quickly overcome costly measurement gates. We then implement a new error model that captures non-unitarity in T1/T2 error channels much faster and with exact accuracy for most physical qubits, demonstrate a chemistry use case, and present a new Clifford+T to Pauli-Based Computing (PBC) transpilation optimization through our simulator.

\end{abstract}

\maketitle

\section{Introduction}
\label{sec:intro}
Quantum computing holds the promise of dramatically accelerating solutions to classically hard problems in areas such as cryptography~\cite{shor1999polynomial, gisin2002quantum}, optimization~\cite{farhi2014quantum, han2002quantum}, quantum chemistry~\cite{georgescu2014quantum, kandala2017hardware}, etc. In recent years, hardware demonstrations have shown an increasing number of qubits while reducing error rates. Quantum devices capable of demonstrating of quantum advantage in certain tasks have been reported~\cite{Google2019, Pan2021, Gao2025}. However, to gain the full potential of quantum computing, the error of quantum systems needs to be carefully managed and corrected~\cite{nielsen_chuang_2010} using QEC codes.

QEC encodes many physical qubits into one logical qubit and actively detects and corrects errors that occur in the quantum systems. QEC lays the foundation for fault-tolerant quantum computing. Early demonstrations of small-scale QEC codes have shown encouraging results~\cite{Anderson2021, Krinner2022, Acharya2023, Bluvstein2024, Acharya2025}, but the path toward mid-term and far-term devices featuring thousands of logical qubits demands significant advances in architecture, control, and simulation~\cite{Gidney2021howtofactorbit}. In particular, designing different QEC schemes while understanding the error threshold and noise performance of logical gate operations, requires very efficient  classical simulation tools. 

\begin{figure}[t]
    \centering
    \includegraphics[width=\linewidth]{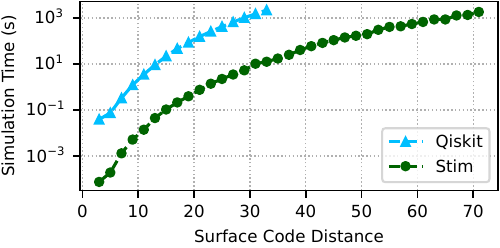}
    \caption{Simulation time of surface codes with distances ranging from 3 to 71, using Qiskit method='stabilizer' and Stim on an AMD EPYC 7763 processor.}
    \label{fig:poor_scaling}
\end{figure}

QEC codes primarily consist of Clifford operations and measurements, which can be efficiently modeled using the stabilizer formalism~\cite{gottesman_heisenberg_1998, nielsen_chuang_2010}. Unlike general quantum circuit simulators that scale exponentially with qubit count, stabilizer simulators scale polynomially, enabling large-scale QEC design verification. Qiskit, in addition to its widely used quantum compilation and emulation features, offers a stabilizer simulation backend that can simulate noisy Clifford circuits if all errors in the noise model are also Clifford errors. This has proven beneficial for debugging near-term algorithms and verifying small to mid-sized QEC codes~\cite{Hayato2024, Bravyi2024, cross2024improvedqldpcsurgerylogical}. Meanwhile, Stim stands out for its meticulous single-threaded CPU optimizations, significantly accelerating large stabilizer circuits through tight memory layouts and specialized bitwise operations~\cite{gidney_stim_2021}. Both Qiskit and Stim highlight the value of specialized stabilizer simulation in quantum computing research, especially for QEC protocols. 

However, as we move forward from the early fault-tolerant era to Megaquop machines~\cite{NISQ_Preskill, Katabarwa2024, Preskill_megaquop}, we expect to expand the simulation capabilities of stabilizer simulators, targeting large-scale multi-logical-qubit code block simulations, with individual patch and multipatch code distances from tens to hundreds. In this regime, the polynomial overhead becomes a challenge for state-of-the-art simulators. Large QEC circuits with repeated rounds of syndrome extraction, fast feedback, and many qubit operations push the limits of CPU-only approaches. Operations like lattice surgery are key for fault tolerance, and have modern tools such as TQEC to enable the construction of enormous fault-tolerant pipelines~\cite{noauthor_tqectqec_2025}.

However, CPU-centric design constraints such as memory bandwidth and low parallelism, place a practical time constraint on actually characterizing errors in these huge architectural constructions. As shown in Figure~\ref{fig:poor_scaling}, while Stim is significantly better than Qiskit, both Qiskit and Stim scale poorly to larger QEC codes. Furthermore, to take advantage of Stim's efficiency via Pauli frames, physical qubit errors must be approximated using Pauli operations, which is an assumption that quickly diverges from reality when considering common non-unitary errors such as qubit relaxation.

Particularly, the volume of stabilizer updates across the stabilizer tableau and scaling faster per-qubit measurement costs quickly overwhelm sequential CPU pipelines, resulting in long simulation times and restricting the scale of practical investigation. In addition, as Stim is highly optimized for shot-based QEC evaluation, other applications relying on Clifford and Pauli operations are difficult to integrate.

To tackle these challenges, we propose a GPU-accelerated stabilizer simulator that exploits the massive parallelism and high-bandwidth memory of modern Graphics Processing Units (GPUs). Originally designed for graphics workloads, GPUs today provide thousands of concurrent threads and efficient memory subsystems, making them an interesting platform for the large matrix-like operations at the core of QEC simulation. Previous efforts on GPU implementation either did not yield performance improvement~\cite{gidney_stim_2021}, or could not perform measurement, which is the key operation to QEC~\cite{osama_parallel_2025}. Stim cited the low arithmetic intensity of bit operations at the core of stabilizer simulation as the reason for GPU under-performance. For measurement, the strict time-order sensitivity and cost of branching posed a barrier that has not been addressed until this work.

Our approach encodes stabilizer states directly on the GPU, orchestrating operations to be applied in parallel with minimal synchronization. Where time-order synchronization is absolutely necessary, extremely efficient CUDA warp-level primitives and optimized block-wide cooperations are employed, allowing us to rapidly clear synchronization and branching bottlenecks. These techniques greatly reduce simulation time as the qubit number and circuit depth increase. This design enables studies of large-scale QEC protocols, including high-distance QEC codes and complex lattice surgery operations which scale impractically on lightly-threaded CPUs. We still provide a CPU version comparable with existing stabilizer simulators as a flexible platform for validating new stabilizer-based algorithms. Both versions support and have been used to develop Pauli-based computing features well beyond direct simulation. As we will show, efficient stabilizer simulation is necessary for large sale QEC, but also has surprising applications for better qubit noise modeling, chemistry simulation, and quantum circuit optimization. This work thus detail four core contributions:
\begin{enumerate}
    \item The first quantum stabilizer simulator to show a performance gain via GPU in Clifford+Measurement simulation. Within a hundreds of qubits, \framework exceeds state-of-the-art simulators in QEC experiments thanks to highly-efficient CUDA warp primitives combined with careful synchronization for accuracy.
    \item Implementation of a new qubit noise model that can be sampled with much more efficiency and accuracy than existing simulators. When $T1$ coherence equals or exceeds $T2$, where most physical qubits reside, our implementation only requires tableau sampling with constant overhead, but has exact accuracy to the physical noise channel. Even when $T2$ is greater, our implementation still requires much fewer tableau samples than other exact methods.
    \item A quantum chemistry use-case for VQE using fast bitwise Pauli string methods available on \framework.
    \item A novel Clifford+T to PBC transpiler using the stabilizer framework. We detail how stabilizer simulation can improve universal circuit architectures by optimizing T-gate count, and show how our method greatly improves the speed of transpilation versus other frameworks.
\end{enumerate}
\section{Background}
\label{sec:background}

\subsection{Clifford Gates}
\label{sec:clifford_background}
At the core of most quantum computing circuits are Pauli operators, which consist of tensor products of Pauli operators $X$, $Y$, $Z$, and identity operator $I$. Pauli operators, combined with phase factors $\{\pm 1, \pm i \}$, form Pauli groups across $n$ qubits. 

The Clifford group comprises all unitary operators that map Pauli operators to other Pauli operators under conjugation. 
Proved by the Gottesman-Knill theorem, quantum circuits restricted to Clifford gates and Pauli measurements can be simulated on a classical computer in polynomial time with respect to the circuit size and number of qubits~\cite{Aaronson2004, gottesman_heisenberg_1998, nielsen_chuang_2010}. Luckily for us, QEC and many other applications are formulated in a way that Clifford gates are sufficient.

\subsection{The Stabilizer formalism}
\label{sec:stabilizer_background}
A stabilizer of a quantum state $\lvert \psi \rangle$ is an operator $\hat{A}$ such that
$\hat{A} \ket{\psi} = +1 \ket{\psi}$.
Although stabilizers can be chosen from broader operator sets, Pauli stabilizers (i.e., tensor products of the Pauli operators) are especially popular due to their simple algebraic structure. A stabilizer group is a commuting (Abelian) subgroup of the $n$-qubit Pauli group whose elements all square to the identity and do not include the negative identity operator $-I$. A state stabilized by a stabilizer group is a stabilizer state. Stabilizer states can also be described by the stabilizers in the stabilizer group, especially the group generators. 

The effect of Clifford group operations on stabilizer states can be simulated by tracking the transformation of the corresponding stabilizer group. Pauli measurement can also be tracked using the following rules~\cite{litinski_game_2019}:
\begin{enumerate}
    \item If the operator $g$ being measured commutes with the stabilizer group $\mathcal{S}$, the operator $g \in \mathcal{S}$. The state will not change after the measurement, while the measurement outcome depends on the phase of the operator $g$ in the group $G$. 
    \item If the operator is not in the stabilizer group $\mathcal{S}$, the measurement outcome can be either $\pm1$. The state after the measurement is stabilized by a new group that contains the new measured operator $\pm g$ depending on the measurement outcome and all other group generators that commute with $g$ in the original stabilizer group $\mathcal{S}$.
\end{enumerate}

A stabilizer $\mathcal{S}$ represents one possible tensor product of Pauli operators that stabilize a state. However, a pure $n$-qubit stabilizer state is stabilized by $n$ commuting and independent Pauli stabilizers. All of these stabilizers can be listed in a table, where rows are the independent stabilizers, and columns are the qubit index.
\begin{equation}
    \left(
\begin{array}{c}
\sigma_{11} \oplus ... \oplus \sigma_{1n} \rightarrow S_1\\
\sigma_{21} \oplus ... \oplus \sigma_{2n} \rightarrow S_2\\
\vdots \\
\sigma_{n1} \oplus ... \oplus \sigma_{nn} \rightarrow S_n\\
\end{array}
\right)
\end{equation}

In order to efficiently track the transformation of Pauli operators, Pauli operators are frequently represented by binary vectors. Each Pauli operator $\sigma$ is represented by two bits, an $x$ portion and a $z$ portion. Pauli $X$ is represented by $(1,0)$, $Z\mapsto(0,1)$, $Y\mapsto(1,1)$, and $I\mapsto(0,0)$. The commutation relation of two Pauli operators represented as $(x_1, z_1)$ and $(x_2, z_2)$ can also be easily computed using
$(x_1 \cdot z_2) \oplus (x_2 \cdot z_1)$,
where $\oplus$ is an xor, and a result of 0 means the two Pauli operators commute.

The key purpose of stabilizers in the context of QEC is that errors can easily knock the state out of the stabilizer subspace of the Hilbert space. In other words, our operator $\hat{A}$, or set of operators, will generate a -1 eigenvalue when it should be stabilizing the state, alerting us to an error.

\subsection{The CHP Formalism}
\label{sec:chp_background}
Aaronson and Gottesman proposed an optimized version of the stabilizer tableau for optimizing measurements, known as the CNOT-Hadamard-Phase (CHP) formalism~\cite{Aaronson2004}. For an $n$-qubit state, instead of a single tableau that only tracks the $n$ stabilizer group generators, in CHP formalism, an extra tableau is added to track the $n$ destabilizers. Specifically, the tableau that are tracked in the CHP formalism can be represented as
\begin{equation}
    \left(
\begin{array}{ccc|ccc|c}
x_{11} & \cdots & x_{1n} & z_{11} & \cdots & z_{1n} & r_1 \\
\vdots & \ddots & \vdots & \vdots & \ddots & \vdots & \vdots \\
x_{n1} & \cdots & x_{nn} & z_{n1} & \cdots & z_{nn} & r_n \\
\hline
x_{(n+1)1} & \cdots & x_{(n+1)n} & z_{(n+1)1} & \cdots & z_{(n+1)n} & r_{n+1} \\
\vdots & \ddots & \vdots & \vdots & \ddots & \vdots & \vdots \\
x_{(2n)1} & \cdots & x_{(2n)n} & z_{(2n)1} & \cdots & z_{(2n)n} & r_{2n}
\end{array}
\right) \nonumber
\end{equation}
where each row represents a stabilizer or destabilizer. The first $n$ rows correspond to the destabilizers, which eliminate the need for $O(n^3)$ Gaussian elimination~\cite{Aaronson2004}, while the last $n$ rows are the stabilizers. The columns $r$ represent the sign bits of each Pauli string (stabilizer). The tableau requires $4n^2+2n$ bits to store all necessary stabilizers and destabilizers, including a phase bit for each, for CHP operations. Thankfully, this means that a full tableau representation of systems with tens of thousands of qubits still only requires space on the order of MB. For comparison, a state vector or density matrix representation of the state will grow on the order $O(2^n)$ and $O(4^n)$ respectively, becoming infeasible with a few dozen qubits~\cite{chen_validating_2025}.

\subsection{Operations on the CHP Tableau}
\label{sec:chp_operations}
Once the data structure of the tableau is established, we want to perform minimal bitwise operations to simulate the evolution of the tableau under Clifford operations. Clifford gates transform Pauli operators to Pauli operators under conjugation. For example, $HXH^\dagger=Z$. Some basic bit transforms on the tableau can be calculated that represent the action of Clifford gates on the stabilizers of a system. Most often, all stabilizers in the tableau will receive the same transformation gate-by-gate, meaning Clifford gate operations can be similar to matrix operations. However, the most time-consuming operations, i.e., measurements and resets, require much more expensive logic.

\begin{table}[h!]
\centering
\begin{tabular}{@{}lll@{}}
\toprule
\textbf{Gate} & \textbf{Binary Tableau Operation~\cite{Aaronson2004}} \\ \midrule
Hadamard ($H$) & Swap $x$ and $z$; $r = r \oplus (x \cdot z)$ \\
Phase ($S$)& $z = z\oplus x$; $r = r \oplus (x \cdot z)$ \\
CNOT ($\mathrm{CX_{c,t}}$) & $x_t = x_t \oplus x_c$; $z_c = z_c \oplus z_t$; \\ & $r =  r\oplus  x_c \cdot z_t$ \\
Measure ($M$) & $ \begin{cases}
  rowsum(i,p)\{\forall  i\mid x_{i,q} = 1 \}(rand)\\
  rowsum(temp,i){\forall  i >n}(determ)\\
\end{cases}$\\
\bottomrule
\end{tabular}
\caption{Clifford gates and their bit operations on the stabilizer tableau. Hadamard, phase and CNOT operations occur on every row of the tableau, at the column that the gate targets. \textit{c} and \textit{t} represent the control and target bit of a CNOT operation. The original \textit{rowsum} algorithm is detailed in~\cite{Aaronson2004} and also provided in Appendix~\ref{app:rowsum} for convenience.}
\end{table}

Other Clifford gates can be derived from this basic gate set and minimized. Performing measurements is sped up by tracking a corresponding destabilizer tableau, but each measurement still requires two distinct steps that branch depending on if the measurement is random or deterministic.

The first step is to search for a stabilizer that does not commute with the measurement. A non-commuting stabilizer indicates a random measurement outcome. In that case, only non-commuting stabilizers are updated. Suppose \textit{p} is the first stabilizer that does not commute with the measurement. \textit{rowsum(i,p)$\{ i \mid x_{q,i} = 1 \}$} (Appendix~\ref{app:rowsum}), updates the phase bit of any non-commuting row \textit{i} using row \textit{p}. The current stabilizer at \textit{p} is then stored in the corresponding destabilizer, and the stabilizer at \textit{p} is projected to the measurement basis, Z in this case. Finally, a random phase bit for \textit{p}, representing the measurement outcome, is chosen.

If all stabilizers commute, the measurement outcome is deterministic. In this case, all stabilizers stay the same, but a temporary bit array of size $n+1$ is used to collect all of the stabilizer updates. \textit{rowsum(temp,i)} $\forall (i>n)$ collects the outcome from every stabilizer in the tableau. Once complete, the phase bit at \textit{temp} is the measurement outcome.

\subsection{Stabilizer Simulator Execution}
\label{sec:execution}
\begin{figure}[t]
\centering
\scalebox{1}{
\begin{tikzpicture}[
  node distance=.3cm and .3cm,
  every node/.style={font=\footnotesize},
  box/.style={rectangle, draw, rounded corners, minimum width=2cm, minimum height=0.5cm, align=center},
  decision/.style={diamond, draw, aspect=2, minimum width=2.5cm, align=center},
  arrow/.style={->, >=Stealth, thick}
  ]

\node[decision] (gateType) {Clifford+M Circuit};

\node[box, below=of gateType, xshift=-2.6cm] (singleQ) {Single Gate(\textit{q})};
\node[box, below=of gateType, xshift=2.6cm] (meas) {Measurement(\textit{q})};
\node[box, below= of gateType] (entangle) {Entangling Gate(\textit{c,t)}};

\node[box, below=of meas] (syncMeas) {Synchronize};

\node[box, below=.3cm of entangle, xshift=-1.5cm] (execC) {Execute Clifford};

\node[box, below=.3cm of syncMeas] (execM) {Warp Reduction};
\node[box, below=.3cm of execM] (wait) {Outcome Decision};

\node[box, below=.3cm of execC] (checkDone) {Circuit done?};
\node[box, below=of checkDone] (done) {Simulated Quantum State};

\draw[arrow] (gateType) -- (singleQ);
\draw[arrow] (gateType) -- (entangle);
\draw[arrow] (gateType) -- (meas);

\draw[arrow] (singleQ) -- (execC);
\draw[arrow] (entangle) -- (execC);

\draw[arrow] (meas) -- (syncMeas);
\draw[arrow] (syncMeas) -- node[right] {M}(execM);

\draw[arrow] (execM) -- (wait);
\draw[arrow] (wait) -- (checkDone);

\draw[arrow] (execC) -- node[right] {C}(checkDone) ;

\draw[arrow] (checkDone) -- node[right] {Yes} (done);

\draw[arrow] (checkDone.west) -- ++(-1.2,0) node[left] {No} |- (gateType.west);

\end{tikzpicture}
}
\caption{Race condition-safe processing of a Clifford circuit into an binary tableau on GPU.}
\label{fig:flowchart}
\end{figure}
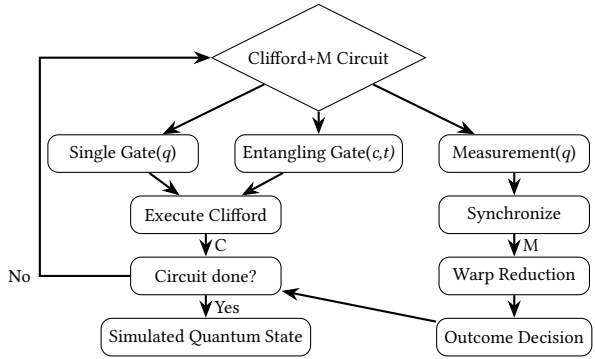

Stabilizer circuits can be split between single-qubit Clifford gates, entangling gates, and measurement gates. The time order of all of these gates must be maintained. For example, phase and Hadamard gates do not commute, entangling gates need to exchange information at a specific point in time, and measurement gates need all relevant stabilizers to be caught up. Time order is more sensitive in the GPU implementation. If multiple threads read and write back to the same location in data, changes from one thread often will not be written back before another thread accesses the data. Careful management of this aspect was required for Section~\ref{sec:stab_design} to provide a speedup without causing inaccuracy due to race conditions. Figure.~\ref{fig:flowchart} provides a high-level overview of how circuits can be safely processed on GPU while maximizing efficiency.

\subsection{QEC Operations}
\label{sec:qec_background}
One of the most important applications using stabilizer simulation is to simulate QEC operations. Specifically, in the syndrome extraction process, the only operations are Clifford gates between syndrome qubits and data qubits, and Pauli-based measurements on those syndrome qubits. When we inject Pauli errors with some probability as a model of actual noise on the physical qubits, the process can be simulated completely in stabilizer formalism. Over some number of shots to a desired accuracy, these noise injections can simulate or approximate different types of physical error.

\begin{figure}[h]
    \centering
    \begin{subfigure}[b]{0.35\textwidth}
        \includegraphics[width=\textwidth]{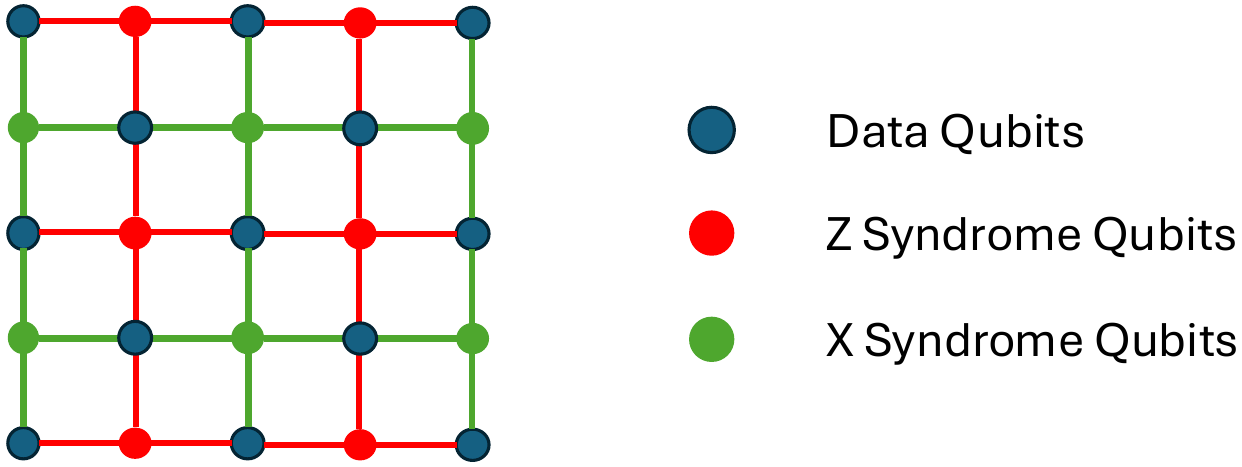}
        \caption{A distance 3 surface code layout}
        \label{fig:layout}
    \end{subfigure}
    \hfill
    \begin{subfigure}[b]{0.275\textwidth}
        \includegraphics[width=\textwidth]{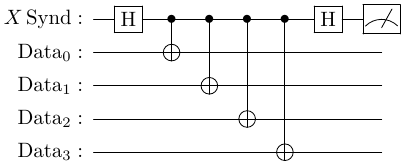}
        \caption{$X$-checks}
        \label{fig:x_checks}
    \end{subfigure} \,
    \begin{subfigure}[b]{0.175\textwidth}
        \includegraphics[width=\textwidth]{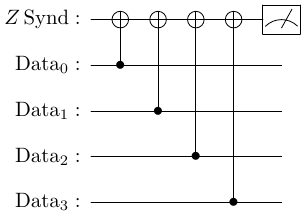}
        \caption{$Z$-checks}
        \label{fig:z_checks}
    \end{subfigure}
    \caption{Planar surface code.}
    \label{fig:surface_code}
\end{figure}

One thoroughly explored QEC code is the planar surface code. Planar surface code is a topological QEC code realized on a two-dimensional grid of physical qubits with open boundary conditions, enabling local stabilizer measurements and scalable fault tolerance. The qubit layout in a distance $5$ surface code patch is shown in Fig.~\ref{fig:layout}, where the data qubits (blue) are the qubits encoding the quantum information, while $X$ ($Z$) syndrome qubits are to extract the syndrome information for quantum error correction. For example, an $X$ ($Z$) syndrome qubit is to measure the stabilizer $\otimes_{j\in \mathcal{N}} X_{j}$ ($\otimes_{j\in \mathcal{N}} Z_{j}$), where $\mathcal{N}$ is the set of neighbor qubits. The syndrome extract circuits for X and Z stabilizers are shown in Fig.~\ref{fig:x_checks} and~\ref{fig:z_checks}, respectively. In a QEC cycle, the X and Z stabilizers of the code patch are extracted, which will be stored and sent to decoders for error correction~\cite{Dennis2002, Fowler2012, Fowler2012_decoder_PRA, Fowler2012_decoder_PRL, Higgott2025sparseblossom}.

\subsection{Stabilizer Noise Modeling}
\label{sec:noise_background}
Density-matrix simulation can accurately capture the full mathematical scope of quantum channels, but requires exponential memory space scaling as O($4^n$). 
For this reason, state-of-the-art stabilizer simulations exclusively permit stochastic Pauli error channels (SPEC) as the method for noise simulations~\cite{gidney_stim_2021, qiskit}, which can be efficiently modeled using the Pauli frame simulation~\cite{gidney_stim_2021}. 
However, non-Pauli error channels can only be approximated using Pauli twirling~\cite{knill_randomized_2007}, which constructs a SPEC by averaging the existing quantum channel conjugated by each Pauli operator.
Pauli twirling is efficient, but inherently loses the correlations of different Pauli bases of the quantum channel that gets twirled. 
Though efficient, Pauli twirling approximation can cause exaggerated logical error rates multiple times worse than reality~\cite{tomita_low-distance_2014, katabarwa_logical_2015}.
In our testing, a single $\ket{+}$ qubit undergoing relaxation had only ~80\% fidelity when Pauli twirled compared to the ground truth when the error duration approached $T1$, which will be further amplified with more error locations and qubits.  
This inaccuracy is only amplified by considering more rounds and qubits. To achieve these effects in stabilizer simulation, 

An exact way of simulating a non-Pauli error channel is to decompose it into stochastic Clifford+Reset gates.
It is not commonly used in QEC investigations for two primary reasons; a) the Pauli frame sampling method, particularly leveraged by Stim for fast noise sampling, is \textit{incompatible} with stochastic non-Pauli noise, and b) the sampling cost is well known to scale \textit{exponentially} with the system size and negativity of the distribution~\cite{bennink_unbiased_2017}.

\subsection{Pauli-Based Computing Transpilation}
\label{sec:pbc_background}
In recent developments of QEC, especially in superconducting qubit based surface code constructions, the limited connectivity between logical patches has led to the proposal of implementing CX gates via lattice surgery. This approach enables straightforward multi-qubit Pauli measurements at the logical level and supports T-gate implementation through multi-patch Pauli measurements combined with magic states.

Pauli-based computing (PBC) is a quantum computational scheme where qubits undergo a sequence of nondestructive eigenvalue measurements on Hermitian Pauli operators—which may be chosen adaptively based on prior measurement outcomes—and the final result is a classical bit derived from polynomial-time processing of the recorded eigenvalues~\cite{PBC}. PBC can readily describe the lattice-surgery-based surface code computation model. Specifically, in~\cite{litinski_game_2019}, a method is introduced to remove all Clifford operations in a quantum circuit by (1) commuting the Clifford gates through the T gates and expanding the T gates into multi-qubit Pauli rotations and (2) absorbing the Clifford operations into measurements to form multi-qubit Pauli measurements. The transformation rule can be derived as
\begin{equation}
\left\lbrace
\begin{array}{l l}
    P \rightarrow P, &  \text{if } [P, P_c] = 0\\
    P \rightarrow i P_c P, &  \text{if } [P, P_c] \neq 0\\
\end{array} 
\right.,
\end{equation}
where $P$ is the Pauli-$\pi/4$ rotation axis (represented as a Pauli operator) or the measurement basis, and $P_c$ is the Pauli operator corresponding to the Clifford rotation angles~\cite{litinski_game_2019}.

\subsection{Pauli Grouping}
\label{sec:pauli_grouping_background}

In quantum algorithms, extracting information from the quantum states prepared by a circuit is crucial, especially in applications such as quantum chemistry, where variational quantum eigensolvers require measuring the expectation value of a Hamiltonian~\cite{Peruzzo2014, tilly_variational_2022, Cerezo2021}. However, quantum hardware natively supports only Pauli-$Z$ measurements on individual qubits. Consequently, evaluating the expectation value of problem Hamiltonians requires decomposition into Pauli strings that can be measured with existing hardware.

To minimize the number of circuit executions shots, a common strategy involves grouping commuting Pauli strings together to measure them simultaneously~\cite{Yen2023}. 
Two popular approaches are Qubit-wise commutation (QWC)~\cite{verteletskyi_measurement_2020} and Group-wise commutation (GC)~\cite{crawford_efficient_2021}. Suppose we have two $n$-qubit Pauli strings $P_1 = \bigotimes_{j=1}^{n} \sigma_{k_j}^{(j)}$ and $P_2 = \bigotimes_{j=1}^{n} \tau_{l_j}^{(j)}$, where $j$ is the index for qubit, $\sigma_{k_j}^{(j)}, \tau_{l_j}^{(j)} \in \{I, X, Y, Z\}$. Under QWC, any two Pauli strings in the same set must satisfy $\forall j$, $[\sigma_{k_j}^{(j)}, \tau_{l_j}^{(j)}] = 0$,
whereas GC only requires $[P_1, P_2] = 0$. Notably, Pauli strings grouped under QWC can be diagonalized using a single layer of single-qubit gates, while GC-based groupings typically require solving for a more complex Clifford circuit to diagonalize the entire set at once.
\section{\framework Design}
\label{sec:stab_design}

We provide two different implementations of the stabilizer simulator, a lightweight CPU version and a parallelized CUDA version. The CPU version implements the tableau as a 2-D grid of independent boolean values for simplicity and quickness. Single boolean values in a 2-D array are intuitive to access and manipulate when extending the simulator for tasks beyond direct simulation. CHP operations are performed sequentially, bit-by-bit on the tableau using loops. While straightforward and fast for small simulations, this approach grows significantly at larger qubit sizes. Every Clifford gate update requires a $\theta(n)$ sweep over the tableau to keep the stabilizers up to date, and measurements can be up to $O(n^2)$ in the worst case. For that reason, we focus on \framework-GPU for our results.

\subsection{The GPU Implementation}
\label{sec:stab_gpu}

For the GPU implementation, the stabilizer tableau built in \framework is stored in global memory and accessed by indexing flattened $x$ and $z$ arrays. Shared memory is avoided for this task since multi-qubit gates and measurements would require reloading and syncing shared memory across blocks. In testing, handling memory between shared blocks meaningfully slowed down circuits with multi-qubit and measurement gates. Instead, each thread is responsible for its own stabilizer row of the tableau for non-measurement operations. Parallelism across rows avoids race conditions for non-measurement gates and allows the tableau to scale quickly for \textit{n}-qubits without additional per-thread overhead.

In contrast to a boolean or packed-bit tableau, which must iterate over a fixed rows for each gate, \framework's GPU design parallelizes row updates, keeping latency low as circuit width increases. Because a standard CHP designed tableau contains $2n$ rows for an $n$-qubit system, the time to simulate a single Clifford gate increases linearly with qubit count in a CPU simulation. Since stabilizers evolve independently under single-qubit Clifford operations, GPU threads can apply gate transformations to each row concurrently, as long as measurement dependencies are accounted for.

Measurement operations cost the majority of simulation time in dynamic QEC circuits where syndrome extraction occurs in many repeated rounds, in agreement with previous works~\cite{gidney_stim_2021}. With that in mind, some key design choices were made to make the most effective use of GPU multi-threading and warp-level primitives.

\subsubsection{Threading}
Threads are responsible for the stabilizers (rows) of the tableau and not the qubits (columns). Stabilizers are generally independent of each other in the tableau instruction set until measurement. In other words, for bulk Clifford gates, each thread accesses memory in the same pattern but unique locations and performs the same bit operations as every other thread. This serves to eliminate the need for synchronization in terms of time-order correctness or any chances of race conditions when rewriting to the tableau.

\subsubsection{Determining the Measurement}
Measurements first require the tableau to be up-to-date to search for a stabilizer that does not commute with the measurement, and that all threads agree on. Then measurements need to update potentially the entire remaining $n^2$ tableau. A few techniques are employed to mitigate the time spent within measurement gates which require costly synchronization bottlenecks.

The search for a stabilizer that does not commute with the measurement starts with a two step block and grid reduction. All blocks find their candidate smallest stabilizer index where x=1 at the gate qubit, for a Z measurement, using $atomicMin$. Only blocks that have found a candidate that is less than the number of rows then submit their candidate to a global reduction. Using atomic operations here is quite cheap; in a typical QEC construction, a thread will only participate in the atomic operations when an error has occurred and the syndrome measurement is no longer deterministic. After those two reductions, a lowest index has been agreed on, with only participation from threads who find anti-commuting stabilizers.

\subsubsection{Warp Primitive Tableau Reduction}
Whether an anti-commuting stabilizer has been found determines the type of measurement outcome. In either case, an inner sum of each stabilizer may need to be done followed by a second reduction to a single value if the measurement is deterministic~\ref{alg:rowsum}. To prevent memory-safe but costly atomic operations from queuing and creating a bottleneck, CUDA warp primitives are heavily relied on. The random measurement begins with a stride over the number of participating warps. This stride is treated as the rows of tableau, since we know every column will participate if a stabilizer row is apart of the measurement update. The second step is to filter out non-participant rows, then stride over the number of lanes in a warp, for each column in each row. Each lane calculates a polynomial which will be a portion of the power to which $i$ is raised when the Pauli operators represented by the first and second row in a $rowsum$ iteration are multiplied. Each lane can also perform its own updates to the tableau in the random case after this polynomial is calculated. This phase is then quickly merged across all warps acting on different columns via $\_\_shfl\_down\_sync$.

Random outcomes are simpler upon inspection, because updates happen on independent stabilizer so separate threads can calculate the phase outcome on each row of the tableau. Deterministic outcomes are more complicated, since the measurement outcome is calculated recursively as rows are iterated, with dependency on the phase outcome of the last row. Furthermore, one scratch row is updated by every participating row in the deterministic measurement, so any way to parallelize this single-index update by multiple rows causes a race condition. Deterministic measurements sweep through the tableau to find the final outcome. This sweep has to keep track of negative, positive, and zero powers as the phase is modified during calculation, even though the CHP formalism guarantees a single-bit outcome phase at the end. To do this, large warp reductions are performed, summing all of the columns of each row, row by row. Each block in \framework contains the maximum number of threads allowable, 1024, to increase access to the same shared memory and maximize where cheaper in-block syncs can be performed, rather than grid-wise syncs. Another consideration is that to perform grid-wise syncs without aborting the GPU kernel, the cooperative thread groups primitives allow only one thread block per streaming-multiprocessor (SM). Therefore, using the maximum allowable threads offers the best occupancy.

As 1024 being divisible exactly twice by the number of threads in a each warp, only two warp primitives are needed to rapidly reduce the sum of 1024 unique polynomials, calculated by each thread, down to only one sum per block. Since this task stays in-block, its natural to utilize shared memory arrays. Finally, the first thread in each block atomically sums with every other block. Most often, blocks will have a 0 internal sum and thus will not participate in this atomic sum. Still, this atomic will only have $n$/1024 participants in the absolute worst case.

Pseudocode~\ref{alg:determ_m} is provided for the two-stage warp reduction to illustrate how the tableau is parsed to maximize the efficiency provided by CUDA warp-level parallelism. Expanding $rowsum$ to a single polynomial helps eliminate thread divergence, aside from functions where only warp or block leaders participate.

\begin{algorithm}
\caption{Phase Polynomial} \label{alg:phase_poly}
\begin{algorithmic}[1]
\Function{$f$}{$x_p, z_p, x_h, z_h$}
    \State \Return $x_p \cdot \big( z_p \cdot (z_h - x_h) + (1 - z_p)\cdot z_h \cdot (2x_h - 1) \big)$
    \Statex \hspace{2.3cm} $+ \; (1 - x_p)\cdot z_p \cdot x_h \cdot (1 - 2z_h)$
\EndFunction
\end{algorithmic}
\end{algorithm}

\begin{algorithm}
\small
\caption{Deterministic-$M$ Warp Reduction} \label{alg:determ_m}
\begin{algorithmic}[1]
\State {$\forall$ rows $\in$ stabilizers:}
\Statex \textbf{--- Stabilizer Reduction ---}
\State $anticommutes \gets x\_arr[k \cdot cols + a]$
\If{$anticommutes$}
    \State $row\_idx \gets row \cdot cols + i$
    \State $thread\_sum \gets 0$

    \If{$i < cols$}
        \Statex \textbf{--- Threaded Sum ---}
        \State $x_r \gets x\_arr[row]$
        \State $z_r \gets z\_arr[row]$
        \State $x_s \gets x\_arr[scratch]$
        \State $z_s \gets z\_arr[scratch]$
        \State $thread\_sum \gets f(x_p, z_p, x_h, z_h)$(Alg~\ref{alg:phase_poly})
        \State $x\_arr[scratch] \gets x_s \oplus x_p$
        \State $z\_arr[scratch] \gets z_s \oplus z_p$

        \Statex \textbf{--- Stage 1: Intra-Warp Reduction ---}
        \State $warp\_sum \gets \texttt{reduce\_add\_sync}(thread\_sum)$
        \If{$block\_lane\_id = 0$}
            \State $\texttt{shared\_mem}[warp\_id] \gets warp\_sum$
        \EndIf
    \EndIf
    \State \texttt{syncthreads()}

    \Statex \textbf{--- Stage 2: Intra-Block Reduction ---}
    \If{$threadIdx.x < 32$}
        \State $block\_sum \gets \texttt{reduce\_add\_sync}($
        \Statex \hspace{1.5em} $\texttt{shared\_mem}[threadIdx.x])$

        \Statex \textbf{--- Global Reduction ---}
        \If{$threadIdx.x = 0 \land block\_sum \neq 0$}
            \State $\texttt{atomicAdd}(global\_sum, block\_sum)$
        \EndIf
    \EndIf

    \State \texttt{grid.sync()}

    \If{$i = 0$}
        \State $r\_arr[scratch\_row] \gets$
        \Statex \hspace{1.5em} $\begin{cases}
            0 & \text{if } (global\_sum + 2r[row] + 2r[scratch]) \bmod 4 = 0 \\
            1 & \text{otherwise}
        \end{cases}$
        \State $global\_sum \gets 0$
    \EndIf

    \State \texttt{grid.sync()}
\EndIf
\end{algorithmic}
\end{algorithm}

\subsubsection{Scheduling Experiment}
\label{sec:schedule_design}
Parallelizing column wise, i.e. assigning other threads to other gates and qubits, effectively shrinks the simulation depth of the circuit, providing another dimension of performance improvement. When parallelizing gate operations, gates should be independent of each other to avoid overwriting tableau data while being used by another thread. Since individual gate operations are so cheap, usually a handful of $xor$ operations, it is easy for threads to jump ahead without synchronization, especially after thousands or millions of operations. This idea lends itself to a scheduler, where multiple independent gates are performed simultaneously. Scheduling at run time seems useful, but did not benefit our experiments. Even the simplest comparisons used to ensure the next gate could safely proceed take longer than the time to directly compute the current gate in the case of a single-pass simulation. By defining flags in the circuit at circuit generation time, it could however be preprocessed into safely independent chunks. With \framework-GPU, the user can call $sim(circuit)$ to perform a gate-wise simulation, or $sim2d(cirucit, chunk\_size)$ to parallelize gates within portions of the circuit where operations are independent from one another. In testing QEC-like circuits, little to no advantage was found using gate-parallelization. QEC runtime is dominated by the time to evaluated mid-circuit measurements, so improving the extremely cheap chunks of non-measurement gates in between had little to no effect. This function is still included in \framework because better suited workloads could utilize parallelized Clifford gates not separated by measurement to speed up simulation time~\cite{osama_parallel_2025}.

\subsection{Pauli Grouping Using \framework}
\label{sec:pauli_group_design}

Pauli grouping is a well used and understood method in quantum chemistry and optimization. Because of this usefulness, and applications to algorithms~\ref{alg:t_sep} and \ref{alg:t_opt} in our T-Gate transpiler, we quickly discuss functions for grouping in the context of binarized stabilizers.

Using $x$ and $z$ as the bits of the $X$ and $Z$ components of the stabilizers (Pauli strings), $(x^{i}_{1} \land z^{i}_{2}) \oplus (x^{i}_{2} \land z^{i}_{1}) = 0$ if they commute at each qubit $i$. Running an xor product can extend the commutation check from qubit-wise to group-wise over the full Pauli string:
\begin{equation}
\bigoplus_{i=1}^n ((x^{i}_{1} \land z^{i}_{2}) \oplus (x^{i}_{2} \land z^{i}_{1})) = 0 
\end{equation}
if $P_1$ and $P_2$ commute. If the binary vectors can be packed into larger data types (e.g. 32, 64-bit words), they can be quickly processed with vectorized bitwise instructions. For group-wise commutation, the xor product reduction can be replaced in favor of a population count on the vectorized results if its an available intrinsic. For example, if 
\begin{equation}
\mathrm{popcount}\left( \left\{ i \in [n] \;\middle|\; [P_1^{(i)}, P_2^{(i)}] \right\} \right) \bmod 2 = 0,
\end{equation}
then $P_{1}$ and $P_{2}$ group-commute.

To evaluate the commutation of a Pauli string \( P \) against an entire stabilizer tableau using matrix operations, the following matrix expression for all stabilizers \( S \) in the tableau is used:
\begin{equation}
(x_P \cdot Z_S) + (z_P \cdot X_S)\mod 2 =0
\end{equation}
if \( S \) commutes with the full tableau. Here, $x_P$ and $z_P$ are the binary row vectors representing the $X$ and $Z$ components of the Pauli string $P$. $X_S$ and $Z_S$ are the binary matrices corresponding to the $X$ and $Z$ components of all stabilizers $S$ in the tableau.
The final sum matrix is an $n \times n$ matrix of the qubit-wise commutation results from $[P, S_{i}]$ where $i$ is each row.

\subsection{Clifford Push Through Transpiler}
\label{sec:push_design}

Transpiling a Clifford+T universal quantum circuit into Pauli measurements with a separate T-layers has great potential to make efficient use of quantum machinery by abstracting away Clifford operations, and minimizing expensive magic state operations. Commuting T-layers can be optimized after separation by removing redundant eighth rotations and pushing them to the Clifford layer as quarter rotation gates.

\subsubsection{Measurement Tableau and T-Tableau Construction}
\label{sec:design:con}
With the target Clifford+T circuit arranged in time order, gates are extracted from the end and added to one of two sets of tableaus. The first tableau is an n-qubit identity tableau of only stabilizers, used for tracking Clifford operations absorbed into the measurement. The second is an empty n-qubit tableau used for tracking the commutation rules of T gates as they are pushed through the circuit. If the gate extracted is a T gate, it is appended to the end of the T tableau as an n-qubit Pauli string with a Z stabilizer at the qubit index of the T gate. If the gate extracted is a Clifford operation, it is applied to the measurement (M) tableau by CHP gate operations. It is also applied to the T tableau by the same rules, but only acts on the T gates which have been appended up to that point. For example, if from the end of the original circuit, no T gates have been extracted yet, then the Clifford gate will have no affect on the T tableau as there are no interactions with the T gates that come before it in time order. Detailed tableau construction psuedocode is provided in Appendix~\ref{alg:t_con}.


\subsubsection{T-Layer Separation}
\label{sec:design:sep}

Once the full circuit has been parsed and M and T tableaus have been created, the T tableau is split into commuting groups as follows. Starting from the end of the T tableau (the last appended stabilizer, i.e. the first T gate in time order) a new empty tableau that we will call $P_{0}$ is made and the stabilizer is appended. Each subsequent stabilizer from the end of the T tableau is checked against the new P tableau using the commutation method discussed in ~\ref{sec:pauli_group_design}. If the subsequent stabilizer commutes with each Pauli string in the first P tableau, it is appended to that tableau. If there is an anti-commutation between the stabilizer being checked and a stabilizer within the tableau, a new tableau ($P_{1}$) is created. For the next stabilizer from the end of T, it is first checked against $P_{0}$. If it commutes with $P_{0}$ it is appended to $P_{0}$. If not, the stabilizer is checked against $P_{1}$ and appended to $P_{1}$ if it commutes. If it anti-commutes, the stabilizer starts a new tableau $P_{2}$. This process continues from the end of T for all stabilizers in T until T is empty. After this process is complete, there will be some number of self-commuting P tableaus (at least one, at most the number of T gates in the original circuit). 

\begin{figure}[h]
    \centering
    \includegraphics[width=0.9\linewidth]{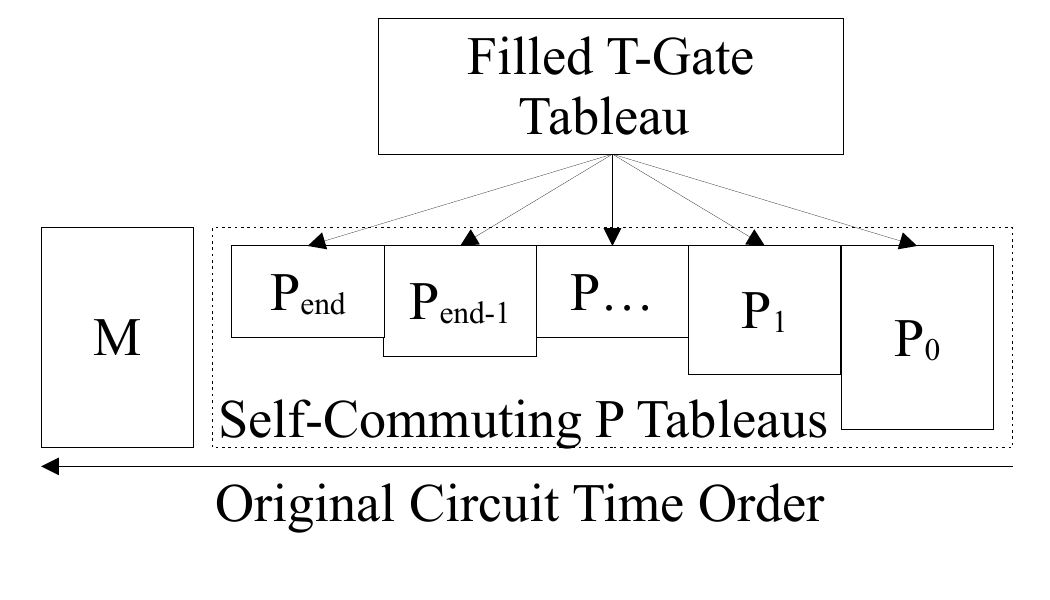}
    \caption{T-Separation. The T-Gate tableau is separated into commuting groups by placing the first T-stabilizer at the end of the original circuit order. If the stabilizer does not commute, it cannot be grouped into the same T-layer, and must be appended to an earlier T-Layer or create its own.}
    \label{fig:t_sep}
\end{figure}

At this point, the basic measurement circuit and T layers have been created. The total number of T-stabilizers is counted to later determine when optimization is done. 
Clifford rotations are now extracted from the P tableaus with the goal of reducing the depth of the T layers. To do this, repeated stabilizers within each tableau are counted, starting from the last P tableau. The P tableaus are in reverse time order, meaning the first entry of $P_{0}$ contains the very last T gate in the original circuit. We want to push Clifford gates to the end, towards the measurement circuit, to absorb any leftover Clifford rotation. For every two occurrences of a stabilizer, a quarter rotation gate can be extracted. Likewise, for every four occurrences of a quarter rotation gate, a full rotation is made, and all eight T stabilizers can be removed without pushing any operations through to the measurement tableau. 

\subsubsection{T-Layer Optimization} 
\label{sec:design:tlayers}

We start by rightward gate pushing from the last P tableau, i.e., the first T gates in original circuit order. Repeat stabilizers in the tableau are counted, removed from the tableau, and the corresponding quarter rotations are then checked  For the second to last P tableau, repeat stabilizers are once again counted and removed, but now need to be commuted through the P tableau to the right before being added to the beginning of the measurement circuit. To do this, one copy of the repeated stabilizer is appended to the next P tableau and checked to see if it commutes with every stabilizer in the tableau. If it does, it is passed through to the next tableau. If not, a $rowsum$(algorithm~\ref{alg:rowsum}) is called on the appended stabilizer with each row of the tableau that does not commute. $rowsum$ accumulates commutation the effect of an anti-commuting Pauli string throughout the stabilizer tableau by summing the contributions of each stabilizer Pauli and corresponding phase bit. The sum modulo 4 of the appended Pauli string respective non-commuting stabilizer becomes the new phase bit $r$ of that stabilizer. $rowsum$ is originally used in the CHP formalism to propagate the effect of measurement on the tableau. To satisfy the commutation rules for pushing a Clifford rotation gate through another arbitrary rotation~\cite{litinski_game_2019}, a phase of $i$ is added to the $rowsum$ product.

\begin{figure}[h]
    \centering
    \includegraphics[width=0.9\linewidth]{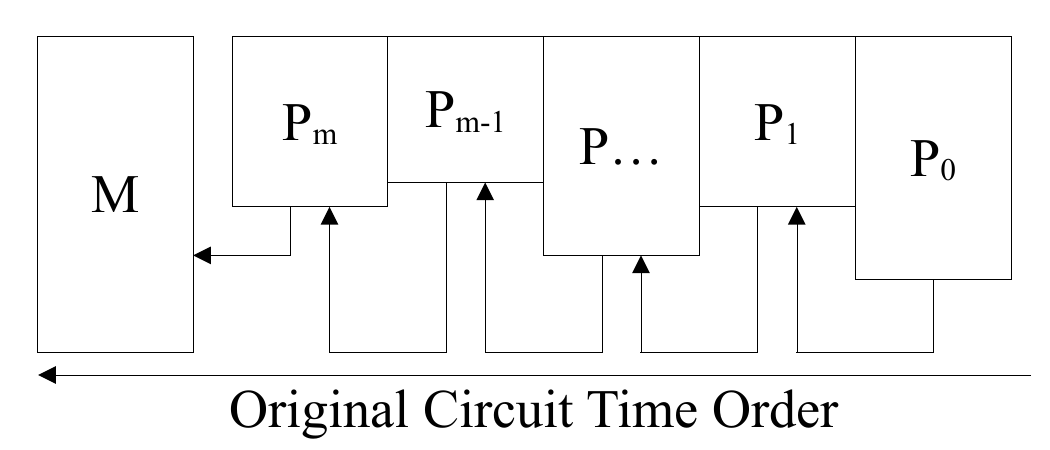}
    \caption{T Optimization. Every two repeating stabilizers are removed from the tableau and pushed left, following the $rowsum+i$ routine for each unique Pauli string.}
    \label{fig:t_opt}
\end{figure}

Once a pass of the T-optimization algorithm is complete, the total T-stabilizers remaining are counted. If no repeat stabilizers were extracted during optimization, the process ends. Otherwise, optimization is repeated to extract new repeat stabilizers created by $rowsum+i$, until the total number of T-stabilizers no longer changes. After optimization, the $P$ tableaus can be recombined to a larger $T$ tableau. The stabilizers of the $T$ tableau are the self-entangled layers of the $\pi/8$ rotation gates that remain. In Figure~\ref{tab:circuit_stats}, every non-Identity Pauli operator in all the layers is counted as a T-rotation, for comparison with the number states needed for T-gates in the original circuit. 

All of our protocols detailed in Section~\ref{sec:push_design} are also rewritten in psuedo-code in Appendix~\ref{app:t_algs} for further reference.

\section{Evaluation}
CPU simulator evaluations were all performed on an AMD EPYC 7502 compute node with a 3.35 GHz boost clock and 256 GB of RAM. GPU evaluations were performed on an NVIDIA H100 SXM node. These platforms were chosen for a stable and consistent test environment versus a laptop or desktop that may have power saving modes and background processes. \framework-GPU used CUDA 12.1.

\subsection{QEC Simulation}
Stabilizer simulation is made feasible in time and size by the relatively slow growth of the stabilizer tableau as qubit count increases compared to other simulation methods. However, measurement operations are $O(n^{2})$ worst-case in the original CHP formalism, one dimension due to every stabilizer potentially being checked, and another from every column participating in $rowsum$ when the row check conditions are met. Measurements greatly bottleneck the time for dynamic circuits, especially QEC codes where many repeated stabilizer measurements are the primary interest for evaluation. 

\begin{figure}[h]
    \centering
    \includegraphics[width=\linewidth]{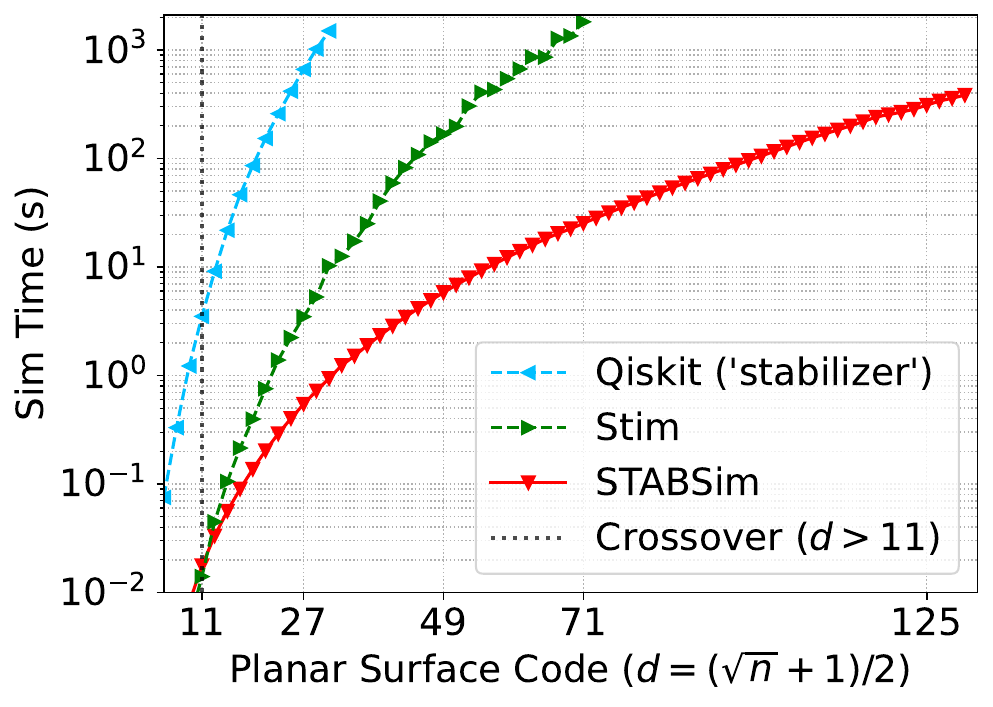}
    \caption{Full tableau simulation time of the planar surface code as code distance increases. Procedural generation highlights faster per-gate simulation times as opposed to deeper targeted compilation. The circuit here follows the example given in Section~\ref{sec:qec_background}.}
    \label{fig:surface_code}
\end{figure}

Stim improves on the deterministic measurement worst case, and, combined with 256-bit SIMD instructions, we find that \framework starts outperforming Stim at distance 11 in the planar code. Beyond distance 11, \framework scales much better than existing simulators as the cost per qubit is almost constant. \framework's scaling is most strongly a function of the number of measurement gates. The primary cost is in synchronization to prepare for warp reduction once all threads have done their computation, which is still quite cheap compared to a $O(n)$ or $O(n^2)$ overhead per gate.

\begin{figure}[h]
    \centering
    \includegraphics[width=\linewidth]{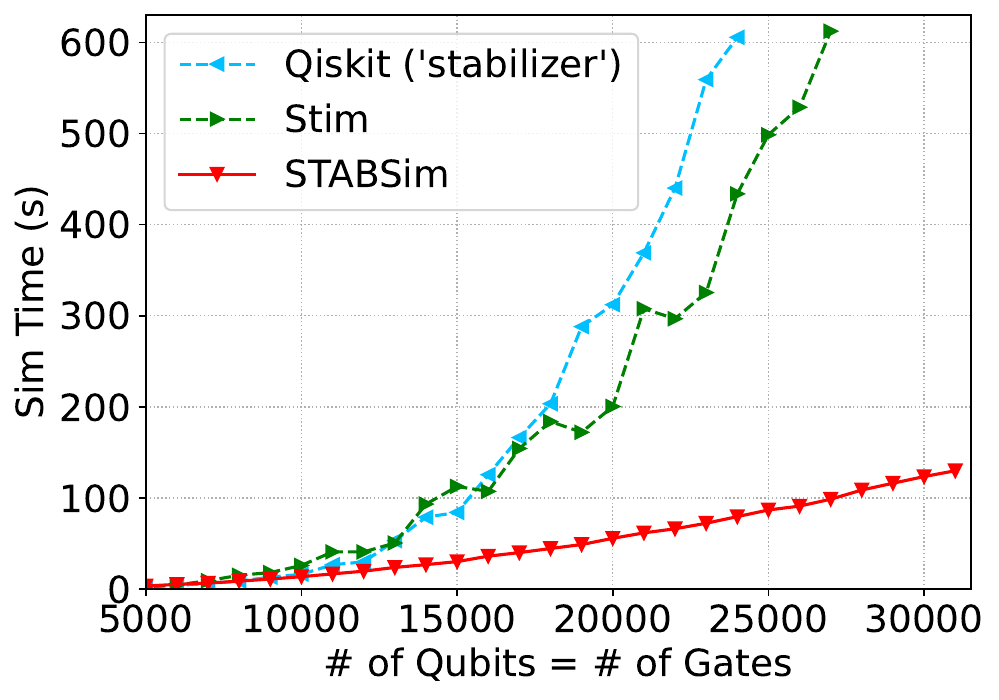}
    \caption{Full tableau simulation of a randomly generated $n$ width circuit with $n$ rounds. This figure highlights performance in scaling with more qubits.}
    \label{fig:random}
\end{figure}

Figure~\ref{fig:random} shows how different simulators scale in a wide circuit random benchmark, where a random gate selected out of H, S, CX, Measurement, and Reset are applied in a number of rounds equal to the number of qubits. This benchmark highlights the polynomial cost of the stabilizer tableau. \framework gets much closer to a linear per-gate cost, versus polynomial gate times in a sequential simulation. Stim and Qiskit both compiled the circuit automatically before simulation, causing deviations depending on how much they were able to simplify each circuit. \framework ran the original circuit directly.

\subsection{Better Noise Modeling}
\label{sec:noise_eval}
\begin{figure*}
    \centering
    \includegraphics[width= \linewidth]{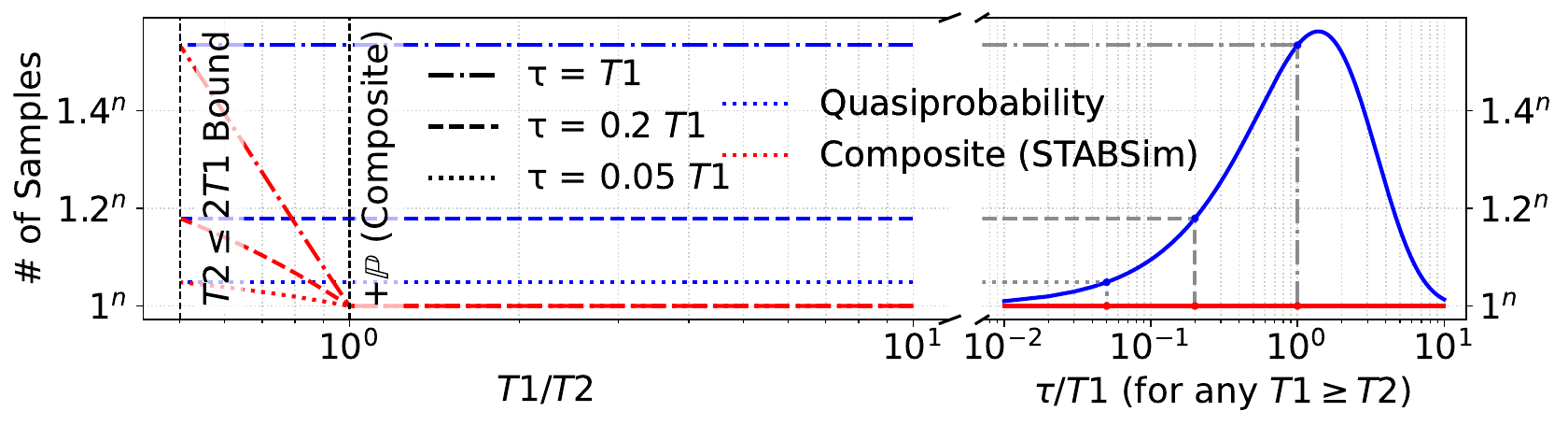}
    \caption{Samples to acheive equal precision, $n$ = qubits. \textbf{Left}: Selected $\mathbf{\tau}$ show how sampling overhead is proportional to the negativity $\mathcal{N}$ which is a function of the ratio of $\mathbf{\tau}$, the error duration, to $\mathbf{T1}$, the relaxation time. The composite probability distribution $\mathbb{P}$ is positive when $T1 \geq T2$, meaning it can be sampled without variance overhead. \textbf{Right}: Sweeping over $\mathbf{\tau}$ in relation to $T1$ visualizes how sampling cost scales with $\mathbf{\tau}$ with prior methods when $T1 \geq T2$.}
    \label{fig:sampling_cost}
\end{figure*}

As discussed in~\ref{sec:noise_background}, existing noise modeling methods generally must either make nonphysical approximations to Pauli channels, or suffer an exponential overhead that grows with qubits (through quasi-probability sampling or density matrix simulation). However, by compositing the qubit relaxation channel with the qubit phase decoherence channel into a single quasi-distribution, 
\[
\mathcal{E}_{phase}(\mathcal{E}_{relaxation}(\rho)),
\]
we implement new a $T1$ and $T2$ noise model that has both constant factor overhead, like SPEC, but also exact accuracy when $T1\geq T2$. Figure~\ref{fig:sampling_cost} shows an analytical evaluation of how the known quasi-probability decomposition for relaxation and decoherence, which can be derived from~\cite{bennink_unbiased_2017}, also implemented in \framework, compares to our composite distribution. 
When $T1\geq T2$, the composite probability is entirely positive, and can be sampled cheaply with Monte Carlo methods. The number of shots required beyond a completely positive baseline distribution is $\propto \Gamma^2$, where $\Gamma$ is the overhead factor of the negative quasi-distribution, defined as $1+2\mathcal{N}$, or $\sum_p{|p|}$ for all probabilities in the quasi-distribution $\mathbb{P}$.

\subsection{Pauli Grouping}
\label{sec:grouping_eval}

To demonstrate a chemistry utility of \framework, we apply \framework to Pauli grouping problems. We focus on grouping the Pauli strings from molecular Hamiltonians, which are widely used in quantum chemistry applications. Hamiltonians were chosen for varying degrees of freedom and measurement gate sizes, while remaining practical to simulate with our state vector simulator. The detailed information about the molecular configuration settings and the Pauli group terms are shown in Table~\ref{tab:circuit_stats}.

To group the Pauli strings, the following steps were employed. First, Pauli terms are sorted up front by the weight of the Pauli term coefficients. Groups of commuting Pauli terms are then greedily constructed starting with the highest weight terms by either QWC or GC. This results in higher weight terms being grouped earlier so sampling resources can be prioritized to get a higher fidelity from more shots on larger terms. Table~\ref{tab:circuit_stats} counts the total number of grouped gates. To verify the correctness of the ground state energy result after grouping, all molecular Hamiltonians were evaluated with a statevector simulator we have developed.

Our tableau-based simulator enables fast and scalable implementation of both strategies with xor product commutation checks. The time spent performing grouping and generating measurement settings was negligible compared to the overall simulation time, especially when considering the time for statevector-based energy verification. VQE in particular still faces challenges in exponential post-processing complexity~\cite{scriva_challenges_2024}, but Pauli grouping is a useful and scalable pre-processing step that can be used with any Pauli-term based quantum algorithm to reduce resource overhead.

\begin{table}[h]
    \centering
    \begin{tabular}{|c|c|c|c|c|c|}     
        \hline
        \textbf{Molecule} & \textbf{$n$} & \textbf{Initial} & \textbf{QWC} & \textbf{GWC}& \textbf{\% Change} \\
        \hline
        H4 & 4 & 3428 & 717 & 275 & 61.65\% \\
        \hline
        LiH & 3 & 756 & 449 & 316 & 29.62\% \\
        \hline
        LiH & 4 & 3428 & 878 & 427 & 51.37\% \\
        \hline
        H6 &  18 & 16196 & 3844 & 1858 & 51.66\% \\
        \hline
        H6 & 22 & 16196 & 3850 & 1946 & 49.45\% \\
        \hline
        BeH2 & 6 & 16196 & 1681 & 749 & 55.44\% \\
        \hline
        BeH2 & 7 & 27124 & 3627 & 1580 & 56.44\% \\
        \hline
    \end{tabular}
    \vspace{2pt}\caption{Gates required for evaluation after grouping small molecular Hamiltonians. Initial basis gates, Qubit-Wise, and Group-Wise gates are shown, along with the \% improvement from QWC to GWC in evaluation gate count. $n$ is the number of qubits/orbitals considered in VQE.}
    \label{tab:circuit_stats}
\end{table}
\subsection{PBC Transpilation}

\begin{figure}[H]
    \centering
    \includegraphics[width=\linewidth]{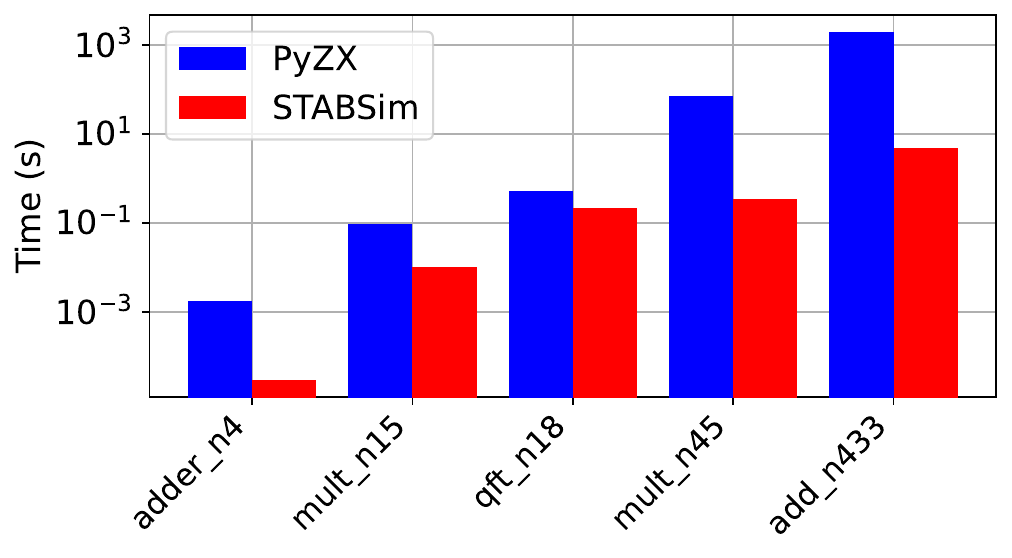}
    \caption{Lower is faster. Transpilation $log_{10}$ time for \framework using the design discussed in Section~\ref{sec:push_design} vs PyZX optimizing on ZX-calculus~\cite{wetering_zx-calculus_2020}. Both converge on the same T-counts.}
    \label{fig:t_comp}
\end{figure}

Like Pauli grouping, the T-Transpiler application relies on fast bitwise commutation checks to be very efficient. Figure~\ref{fig:t_comp} shows the total time \framework took to perform Algorithms~\ref{alg:t_con},~\ref{alg:t_sep}, and~\ref{alg:t_opt}. The purpose of the T-transpiler is to allow for easier scheduling of T-gate factories in quantum architectures by a) lowering the number of T layers, and b) providing an equal cost for each layer. PyZX contains the same functionality, but converts the circuit to a ZX-calculus representation that can then be minimized~\cite{kissinger_reducing_2020}. For all of the circuits tested, both \framework and PyZX converged on the same T-optimization factor. The difference in time is how long it took to arrive at that factor.

Most of the substantial time gain found with \framework is an improvement in the optimization step. Processing is primarily an issue of circuit size, and does not require commutation checks or other applications that can be sped up by bitwise methods. A complete table of optimized Clifford+T to PBC circuit statistics is provided in Appendix~\ref{app:opt_results}. The T-stabilizers remaining after optimization comprise most of the remaining circuit, apart from the now-merged and rotated measurement operations at the end. Each stabilizer in the T-tableau after the full processing is an entangled layer of $\pi/8$ rotational gates in the X, Y, or Z basis, given by the Pauli operator on each qubit. A phase bit of 1 in the stabilizer means the basis rotation in that layer is negative. The circuits with more trivial T-States tended to converge quickly and eliminate many T-layers, which reduced the size and frequency of loops in Algorithm~\ref{alg:t_opt}, where commutation checks become a bottleneck.
\section{Related Work}
Many projects have been developed on the concept of simulating a quantum system from its stabilizers since CHP was proposed by Aaronson and Gottesman~\cite{Aaronson2004}. Cirq, Qiskit, and other projects often implement stabilizer simulators as functions in their libraries~\cite{cirq_developers_2021, qiskit}. More recently, the Stim project made large gains in QEC sampling via meticulous CPU-based optimization and Pauli frame simulation alongside the stabilizer tableau~\cite{gidney_stim_2021}. Those tools are the most comparable to \framework, as we aim to scale up and expand the functions they perform.

Cuda-QEC leverages GPU for accelerated decoding of error syndromes~\cite{cudaqx}. However, it does not provide a stabilizer simulator for characterizing large error correcting codes. Rather, Cuda-QEC focuses on parallelism for decoding syndromes, which much be provided by a simulator. Qua-SARQ performs GPU accelerated simulation of Clifford gates on a stabilizer tableau, to replace the work of Stim in the equivalence checking applications\cite{osama_parallel_2025}. Crucially, Qua-SARQ only simulates Clifford gates, which, given the similarity to large binary matrix operations, are a portion of stabilizer simulation that are straightforward to parallelize. Qua-SARQ does not address measurement gates, which turn out to be the most costly portion of simulation and the crux of syndrome extraction in QEC~\cite{gidney_stim_2021}.

Pauli measurements and Pauli-Based Computation (PBC) are discussed significantly by Litinski in the context of surface code operations for fault-tolerance~\cite{litinski_game_2019}. Multiple works explore PBC compilation using various methods, including a standalone Python compiler using space/time tradeoffs with auxiliary qubits, and a ZX calculus method implemented in PyZX~\cite{peres_quantum_2023, kissinger_reducing_2020}. Our work is the first to formulate PBC in the context of a stabilizer tableau, where highly efficient bitwise stabilizer operations are exploited for much faster compilation than existing methods.
\section{Conclusion}
Large stabilizer simulations become more and more of interest as quantum computing moves towards intermediate scale and eventual fault tolerance. Applications benefiting from large scale stabilizer-based computing can be found in higher-distance error correcting codes, quantum verification, limited Clifford+T circuits, heterogeneous codes, and more~\cite{bravyi_high-threshold_2024, fowler_surface_2012, bravyi_improved_2016,stein_architectures_2024}. \framework shows promise in evaluating growing QEC codes or other large circuits in reasonable amounts of time by scaling better than any existing stabilizer simulator. We furthermore show how the extended capabilities of \framework directly benefit other domains beyond direct simulation. 

Work in these areas has also inspired ideas for future directions. For example, application of our Clifford simulator to other non-Clifford channels decomposed into quasi-probability distributions is interesting, as it could explore a new regime of practical simulation via rapid Monte Carlo sampling. GPU has the potential to also perform parallel sampling, if memory costs can remain practical. Investigation beyond the scope of this work could provide better guidance on which error sampling methods are best to use in different scenarios.

\section*{Acknowledgment}
This research was supported by the Quantum Algorithms and Architecture for Domain Science Initiative (QuAADS), under the Laboratory Directed Research and Development (LDRD) Program at Pacific Northwest National Laboratory (PNNL). M. Wang and A. Li were supported by the U.S. Department of Energy, Office of Science, National Quantum Information Science Research Centers, Quantum Science Center (QSC). This research used resources of the Oak Ridge Leadership Computing Facility, which is a DOE Office of Science User Facility supported under Contract DE-AC05-00OR22725. This research used resources of the National Energy Research Scientific Computing Center (NERSC), a U.S. Department of Energy Office of Science User Facility located at Lawrence Berkeley National Laboratory, operated under Contract No. DE-AC02-05CH11231. The Pacific Northwest National Laboratory is operated by Battelle for the U.S. Department of Energy under Contract DE-AC05-76RL01830.

\bibliographystyle{ACM-Reference-Format}
\bibliography{stabsim}


\begin{thebibliography}{56}


\ifx \showCODEN    \undefined \def \showCODEN     #1{\unskip}     \fi
\ifx \showDOI      \undefined \def \showDOI       #1{#1}\fi
\ifx \showISBNx    \undefined \def \showISBNx     #1{\unskip}     \fi
\ifx \showISBNxiii \undefined \def \showISBNxiii  #1{\unskip}     \fi
\ifx \showISSN     \undefined \def \showISSN      #1{\unskip}     \fi
\ifx \showLCCN     \undefined \def \showLCCN      #1{\unskip}     \fi
\ifx \shownote     \undefined \def \shownote      #1{#1}          \fi
\ifx \showarticletitle \undefined \def \showarticletitle #1{#1}   \fi
\ifx \showURL      \undefined \def \showURL       {\relax}        \fi
\providecommand\bibfield[2]{#2}
\providecommand\bibinfo[2]{#2}
\providecommand\natexlab[1]{#1}
\providecommand\showeprint[2][]{arXiv:#2}

\bibitem[noa(2025)]%
        {noauthor_tqectqec_2025}
 \bibinfo{year}{2025}\natexlab{}.
\newblock \bibinfo{title}{tqec/tqec}.
\newblock
\newblock
\urldef\tempurl%
\url{https://github.com/tqec/tqec}
\showURL{%
\tempurl}
\newblock
\shownote{original-date: 2023-10-19T16:33:34Z}.


\bibitem[Aaronson and Gottesman(2004)]%
        {Aaronson2004}
\bibfield{author}{\bibinfo{person}{Scott Aaronson} {and} \bibinfo{person}{Daniel Gottesman}.} \bibinfo{year}{2004}\natexlab{}.
\newblock \showarticletitle{Improved simulation of stabilizer circuits}.
\newblock \bibinfo{journal}{\emph{Phys. Rev. A}}  \bibinfo{volume}{70} (\bibinfo{date}{Nov} \bibinfo{year}{2004}), \bibinfo{pages}{052328}.
\newblock
Issue 5.
\urldef\tempurl%
\url{https://doi.org/10.1103/PhysRevA.70.052328}
\showDOI{\tempurl}


\bibitem[Acharya et~al\mbox{.}(2025)]%
        {Acharya2025}
\bibfield{author}{\bibinfo{person}{Rajeev Acharya}, \bibinfo{person}{Dmitry~A. Abanin}, \bibinfo{person}{Laleh Aghababaie-Beni}, \bibinfo{person}{Igor Aleiner}, \bibinfo{person}{Trond~I. Andersen}, \bibinfo{person}{Markus Ansmann}, \bibinfo{person}{Frank Arute}, \bibinfo{person}{Kunal Arya}, \bibinfo{person}{Abraham Asfaw}, \bibinfo{person}{Nikita Astrakhantsev}, \bibinfo{person}{Juan Atalaya}, \bibinfo{person}{Ryan Babbush}, \bibinfo{person}{Dave Bacon}, \bibinfo{person}{Brian Ballard}, \bibinfo{person}{Joseph~C. Bardin}, \bibinfo{person}{Johannes Bausch}, \bibinfo{person}{Andreas Bengtsson}, \bibinfo{person}{Alexander Bilmes}, \bibinfo{person}{Sam Blackwell}, \bibinfo{person}{Sergio Boixo}, \bibinfo{person}{Gina Bortoli}, \bibinfo{person}{Alexandre Bourassa}, \bibinfo{person}{Jenna Bovaird}, \bibinfo{person}{Leon Brill}, \bibinfo{person}{Michael Broughton}, \bibinfo{person}{David~A. Browne}, \bibinfo{person}{Brett Buchea}, \bibinfo{person}{Bob~B. Buckley}, \bibinfo{person}{David~A. Buell},
  \bibinfo{person}{Tim Burger}, \bibinfo{person}{Brian Burkett}, \bibinfo{person}{Nicholas Bushnell}, \bibinfo{person}{Anthony Cabrera}, \bibinfo{person}{Juan Campero}, \bibinfo{person}{Hung-Shen Chang}, \bibinfo{person}{Yu Chen}, \bibinfo{person}{Zijun Chen}, \bibinfo{person}{Ben Chiaro}, \bibinfo{person}{Desmond Chik}, \bibinfo{person}{Charina Chou}, \bibinfo{person}{Jahan Claes}, \bibinfo{person}{Agnetta~Y. Cleland}, \bibinfo{person}{Josh Cogan}, \bibinfo{person}{Roberto Collins}, \bibinfo{person}{Paul Conner}, \bibinfo{person}{William Courtney}, \bibinfo{person}{Alexander~L. Crook}, \bibinfo{person}{Ben Curtin}, \bibinfo{person}{Sayan Das}, \bibinfo{person}{Alex Davies}, \bibinfo{person}{Laura De~Lorenzo}, \bibinfo{person}{Dripto~M. Debroy}, \bibinfo{person}{Sean Demura}, \bibinfo{person}{Michel Devoret}, \bibinfo{person}{Agustin Di~Paolo}, \bibinfo{person}{Paul Donohoe}, \bibinfo{person}{Ilya Drozdov}, \bibinfo{person}{Andrew Dunsworth}, \bibinfo{person}{Clint Earle}, \bibinfo{person}{Thomas Edlich},
  \bibinfo{person}{Alec Eickbusch}, \bibinfo{person}{Aviv~Moshe Elbag}, \bibinfo{person}{Mahmoud Elzouka}, \bibinfo{person}{Catherine Erickson}, \bibinfo{person}{Lara Faoro}, \bibinfo{person}{Edward Farhi}, \bibinfo{person}{Vinicius~S. Ferreira}, \bibinfo{person}{Leslie~Flores Burgos}, \bibinfo{person}{Ebrahim Forati}, \bibinfo{person}{Austin~G. Fowler}, \bibinfo{person}{Brooks Foxen}, \bibinfo{person}{Suhas Ganjam}, \bibinfo{person}{Gonzalo Garcia}, \bibinfo{person}{Robert Gasca}, \bibinfo{person}{{\'E}lie Genois}, \bibinfo{person}{William Giang}, \bibinfo{person}{Craig Gidney}, \bibinfo{person}{Dar Gilboa}, \bibinfo{person}{Raja Gosula}, \bibinfo{person}{Alejandro~Grajales Dau}, \bibinfo{person}{Dietrich Graumann}, \bibinfo{person}{Alex Greene}, \bibinfo{person}{Jonathan~A. Gross}, \bibinfo{person}{Steve Habegger}, \bibinfo{person}{John Hall}, \bibinfo{person}{Michael~C. Hamilton}, \bibinfo{person}{Monica Hansen}, \bibinfo{person}{Matthew~P. Harrigan}, \bibinfo{person}{Sean~D. Harrington},
  \bibinfo{person}{Francisco J.~H. Heras}, \bibinfo{person}{Stephen Heslin}, \bibinfo{person}{Paula Heu}, \bibinfo{person}{Oscar Higgott}, \bibinfo{person}{Gordon Hill}, \bibinfo{person}{Jeremy Hilton}, \bibinfo{person}{George Holland}, \bibinfo{person}{Sabrina Hong}, \bibinfo{person}{Hsin-Yuan Huang}, \bibinfo{person}{Ashley Huff}, \bibinfo{person}{William~J. Huggins}, \bibinfo{person}{Lev~B. Ioffe}, \bibinfo{person}{Sergei~V. Isakov}, \bibinfo{person}{Justin Iveland}, \bibinfo{person}{Evan Jeffrey}, \bibinfo{person}{Zhang Jiang}, \bibinfo{person}{Cody Jones}, \bibinfo{person}{Stephen Jordan}, \bibinfo{person}{Chaitali Joshi}, \bibinfo{person}{Pavol Juhas}, \bibinfo{person}{Dvir Kafri}, \bibinfo{person}{Hui Kang}, \bibinfo{person}{Amir~H. Karamlou}, \bibinfo{person}{Kostyantyn Kechedzhi}, \bibinfo{person}{Julian Kelly}, \bibinfo{person}{Trupti Khaire}, \bibinfo{person}{Tanuj Khattar}, \bibinfo{person}{Mostafa Khezri}, \bibinfo{person}{Seon Kim}, \bibinfo{person}{Paul~V. Klimov}, \bibinfo{person}{Andrey~R.
  Klots}, \bibinfo{person}{Bryce Kobrin}, \bibinfo{person}{Pushmeet Kohli}, \bibinfo{person}{Alexander~N. Korotkov}, \bibinfo{person}{Fedor Kostritsa}, \bibinfo{person}{Robin Kothari}, \bibinfo{person}{Borislav Kozlovskii}, \bibinfo{person}{John~Mark Kreikebaum}, \bibinfo{person}{Vladislav~D. Kurilovich}, \bibinfo{person}{Nathan Lacroix}, \bibinfo{person}{David Landhuis}, \bibinfo{person}{Tiano Lange-Dei}, \bibinfo{person}{Brandon~W. Langley}, \bibinfo{person}{Pavel Laptev}, \bibinfo{person}{Kim-Ming Lau}, \bibinfo{person}{Lo{\"\i}ck Le~Guevel}, \bibinfo{person}{Justin Ledford}, \bibinfo{person}{Joonho Lee}, \bibinfo{person}{Kenny Lee}, \bibinfo{person}{Yuri~D. Lensky}, \bibinfo{person}{Shannon Leon}, \bibinfo{person}{Brian~J. Lester}, \bibinfo{person}{Wing~Yan Li}, \bibinfo{person}{Yin Li}, \bibinfo{person}{Alexander~T. Lill}, \bibinfo{person}{Wayne Liu}, \bibinfo{person}{William~P. Livingston}, \bibinfo{person}{Aditya Locharla}, \bibinfo{person}{Erik Lucero}, \bibinfo{person}{Daniel Lundahl},
  \bibinfo{person}{Aaron Lunt}, \bibinfo{person}{Sid Madhuk}, \bibinfo{person}{Fionn~D. Malone}, \bibinfo{person}{Ashley Maloney}, \bibinfo{person}{Salvatore Mandr{\`a}}, \bibinfo{person}{James Manyika}, \bibinfo{person}{Leigh~S. Martin}, \bibinfo{person}{Orion Martin}, \bibinfo{person}{Steven Martin}, \bibinfo{person}{Cameron Maxfield}, \bibinfo{person}{Jarrod~R. McClean}, \bibinfo{person}{Matt McEwen}, \bibinfo{person}{Seneca Meeks}, \bibinfo{person}{Anthony Megrant}, \bibinfo{person}{Xiao Mi}, \bibinfo{person}{Kevin~C. Miao}, \bibinfo{person}{Amanda Mieszala}, \bibinfo{person}{Reza Molavi}, \bibinfo{person}{Sebastian Molina}, \bibinfo{person}{Shirin Montazeri}, \bibinfo{person}{Alexis Morvan}, \bibinfo{person}{Ramis Movassagh}, \bibinfo{person}{Wojciech Mruczkiewicz}, \bibinfo{person}{Ofer Naaman}, \bibinfo{person}{Matthew Neeley}, \bibinfo{person}{Charles Neill}, \bibinfo{person}{Ani Nersisyan}, \bibinfo{person}{Hartmut Neven}, \bibinfo{person}{Michael Newman}, \bibinfo{person}{Jiun~How Ng},
  \bibinfo{person}{Anthony Nguyen}, \bibinfo{person}{Murray Nguyen}, \bibinfo{person}{Chia-Hung Ni}, \bibinfo{person}{Murphy~Yuezhen Niu}, \bibinfo{person}{Thomas~E. O'Brien}, \bibinfo{person}{William~D. Oliver}, \bibinfo{person}{Alex Opremcak}, \bibinfo{person}{Kristoffer Ottosson}, \bibinfo{person}{Andre Petukhov}, \bibinfo{person}{Alex Pizzuto}, \bibinfo{person}{John Platt}, \bibinfo{person}{Rebecca Potter}, \bibinfo{person}{Orion Pritchard}, \bibinfo{person}{Leonid~P. Pryadko}, \bibinfo{person}{Chris Quintana}, \bibinfo{person}{Ganesh Ramachandran}, \bibinfo{person}{Matthew~J. Reagor}, \bibinfo{person}{John Redding}, \bibinfo{person}{David~M. Rhodes}, \bibinfo{person}{Gabrielle Roberts}, \bibinfo{person}{Eliott Rosenberg}, \bibinfo{person}{Emma Rosenfeld}, \bibinfo{person}{Pedram Roushan}, \bibinfo{person}{Nicholas~C. Rubin}, \bibinfo{person}{Negar Saei}, \bibinfo{person}{Daniel Sank}, \bibinfo{person}{Kannan Sankaragomathi}, \bibinfo{person}{Kevin~J. Satzinger}, \bibinfo{person}{Henry~F. Schurkus},
  \bibinfo{person}{Christopher Schuster}, \bibinfo{person}{Andrew~W. Senior}, \bibinfo{person}{Michael~J. Shearn}, \bibinfo{person}{Aaron Shorter}, \bibinfo{person}{Noah Shutty}, \bibinfo{person}{Vladimir Shvarts}, \bibinfo{person}{Shraddha Singh}, \bibinfo{person}{Volodymyr Sivak}, \bibinfo{person}{Jindra Skruzny}, \bibinfo{person}{Spencer Small}, \bibinfo{person}{Vadim Smelyanskiy}, \bibinfo{person}{W.~Clarke Smith}, \bibinfo{person}{Rolando~D. Somma}, \bibinfo{person}{Sofia Springer}, \bibinfo{person}{George Sterling}, \bibinfo{person}{Doug Strain}, \bibinfo{person}{Jordan Suchard}, \bibinfo{person}{Aaron Szasz}, \bibinfo{person}{Alex Sztein}, \bibinfo{person}{Douglas Thor}, \bibinfo{person}{Alfredo Torres}, \bibinfo{person}{M.~Mert Torunbalci}, \bibinfo{person}{Abeer Vaishnav}, \bibinfo{person}{Justin Vargas}, \bibinfo{person}{Sergey Vdovichev}, \bibinfo{person}{Guifre Vidal}, \bibinfo{person}{Benjamin Villalonga}, \bibinfo{person}{Catherine~Vollgraff Heidweiller}, \bibinfo{person}{Steven Waltman},
  \bibinfo{person}{Shannon~X. Wang}, \bibinfo{person}{Brayden Ware}, \bibinfo{person}{Kate Weber}, \bibinfo{person}{Travis Weidel}, \bibinfo{person}{Theodore White}, \bibinfo{person}{Kristi Wong}, \bibinfo{person}{Bryan W.~K. Woo}, \bibinfo{person}{Cheng Xing}, \bibinfo{person}{Z.~Jamie Yao}, \bibinfo{person}{Ping Yeh}, \bibinfo{person}{Bicheng Ying}, \bibinfo{person}{Juhwan Yoo}, \bibinfo{person}{Noureldin Yosri}, \bibinfo{person}{Grayson Young}, \bibinfo{person}{Adam Zalcman}, \bibinfo{person}{Yaxing Zhang}, \bibinfo{person}{Ningfeng Zhu}, \bibinfo{person}{Nicholas Zobrist}, \bibinfo{person}{Google~Quantum AI}, {and} \bibinfo{person}{Collaborators}.} \bibinfo{year}{2025}\natexlab{}.
\newblock \showarticletitle{Quantum error correction below the surface code threshold}.
\newblock \bibinfo{journal}{\emph{Nature}} \bibinfo{volume}{638}, \bibinfo{number}{8052} (\bibinfo{year}{2025}), \bibinfo{pages}{920--926}.
\newblock
\showISBNx{1476-4687}
\urldef\tempurl%
\url{https://doi.org/10.1038/s41586-024-08449-y}
\showDOI{\tempurl}


\bibitem[Acharya et~al\mbox{.}(2023)]%
        {Acharya2023}
\bibfield{author}{\bibinfo{person}{Rajeev Acharya}, \bibinfo{person}{Igor Aleiner}, \bibinfo{person}{Richard Allen}, \bibinfo{person}{Trond~I. Andersen}, \bibinfo{person}{Markus Ansmann}, \bibinfo{person}{Frank Arute}, \bibinfo{person}{Kunal Arya}, \bibinfo{person}{Abraham Asfaw}, \bibinfo{person}{Juan Atalaya}, \bibinfo{person}{Ryan Babbush}, \bibinfo{person}{Dave Bacon}, \bibinfo{person}{Joseph~C. Bardin}, \bibinfo{person}{Joao Basso}, \bibinfo{person}{Andreas Bengtsson}, \bibinfo{person}{Sergio Boixo}, \bibinfo{person}{Gina Bortoli}, \bibinfo{person}{Alexandre Bourassa}, \bibinfo{person}{Jenna Bovaird}, \bibinfo{person}{Leon Brill}, \bibinfo{person}{Michael Broughton}, \bibinfo{person}{Bob~B. Buckley}, \bibinfo{person}{David~A. Buell}, \bibinfo{person}{Tim Burger}, \bibinfo{person}{Brian Burkett}, \bibinfo{person}{Nicholas Bushnell}, \bibinfo{person}{Yu Chen}, \bibinfo{person}{Zijun Chen}, \bibinfo{person}{Ben Chiaro}, \bibinfo{person}{Josh Cogan}, \bibinfo{person}{Roberto Collins}, \bibinfo{person}{Paul
  Conner}, \bibinfo{person}{William Courtney}, \bibinfo{person}{Alexander~L. Crook}, \bibinfo{person}{Ben Curtin}, \bibinfo{person}{Dripto~M. Debroy}, \bibinfo{person}{Alexander Del Toro~Barba}, \bibinfo{person}{Sean Demura}, \bibinfo{person}{Andrew Dunsworth}, \bibinfo{person}{Daniel Eppens}, \bibinfo{person}{Catherine Erickson}, \bibinfo{person}{Lara Faoro}, \bibinfo{person}{Edward Farhi}, \bibinfo{person}{Reza Fatemi}, \bibinfo{person}{Leslie Flores~Burgos}, \bibinfo{person}{Ebrahim Forati}, \bibinfo{person}{Austin~G. Fowler}, \bibinfo{person}{Brooks Foxen}, \bibinfo{person}{William Giang}, \bibinfo{person}{Craig Gidney}, \bibinfo{person}{Dar Gilboa}, \bibinfo{person}{Marissa Giustina}, \bibinfo{person}{Alejandro Grajales~Dau}, \bibinfo{person}{Jonathan~A. Gross}, \bibinfo{person}{Steve Habegger}, \bibinfo{person}{Michael~C. Hamilton}, \bibinfo{person}{Matthew~P. Harrigan}, \bibinfo{person}{Sean~D. Harrington}, \bibinfo{person}{Oscar Higgott}, \bibinfo{person}{Jeremy Hilton}, \bibinfo{person}{Markus
  Hoffmann}, \bibinfo{person}{Sabrina Hong}, \bibinfo{person}{Trent Huang}, \bibinfo{person}{Ashley Huff}, \bibinfo{person}{William~J. Huggins}, \bibinfo{person}{Lev~B. Ioffe}, \bibinfo{person}{Sergei~V. Isakov}, \bibinfo{person}{Justin Iveland}, \bibinfo{person}{Evan Jeffrey}, \bibinfo{person}{Zhang Jiang}, \bibinfo{person}{Cody Jones}, \bibinfo{person}{Pavol Juhas}, \bibinfo{person}{Dvir Kafri}, \bibinfo{person}{Kostyantyn Kechedzhi}, \bibinfo{person}{Julian Kelly}, \bibinfo{person}{Tanuj Khattar}, \bibinfo{person}{Mostafa Khezri}, \bibinfo{person}{M{\'a}ria Kieferov{\'a}}, \bibinfo{person}{Seon Kim}, \bibinfo{person}{Alexei Kitaev}, \bibinfo{person}{Paul~V. Klimov}, \bibinfo{person}{Andrey~R. Klots}, \bibinfo{person}{Alexander~N. Korotkov}, \bibinfo{person}{Fedor Kostritsa}, \bibinfo{person}{John~Mark Kreikebaum}, \bibinfo{person}{David Landhuis}, \bibinfo{person}{Pavel Laptev}, \bibinfo{person}{Kim-Ming Lau}, \bibinfo{person}{Lily Laws}, \bibinfo{person}{Joonho Lee}, \bibinfo{person}{Kenny Lee},
  \bibinfo{person}{Brian~J. Lester}, \bibinfo{person}{Alexander Lill}, \bibinfo{person}{Wayne Liu}, \bibinfo{person}{Aditya Locharla}, \bibinfo{person}{Erik Lucero}, \bibinfo{person}{Fionn~D. Malone}, \bibinfo{person}{Jeffrey Marshall}, \bibinfo{person}{Orion Martin}, \bibinfo{person}{Jarrod~R. McClean}, \bibinfo{person}{Trevor McCourt}, \bibinfo{person}{Matt McEwen}, \bibinfo{person}{Anthony Megrant}, \bibinfo{person}{Bernardo Meurer~Costa}, \bibinfo{person}{Xiao Mi}, \bibinfo{person}{Kevin~C. Miao}, \bibinfo{person}{Masoud Mohseni}, \bibinfo{person}{Shirin Montazeri}, \bibinfo{person}{Alexis Morvan}, \bibinfo{person}{Emily Mount}, \bibinfo{person}{Wojciech Mruczkiewicz}, \bibinfo{person}{Ofer Naaman}, \bibinfo{person}{Matthew Neeley}, \bibinfo{person}{Charles Neill}, \bibinfo{person}{Ani Nersisyan}, \bibinfo{person}{Hartmut Neven}, \bibinfo{person}{Michael Newman}, \bibinfo{person}{Jiun~How Ng}, \bibinfo{person}{Anthony Nguyen}, \bibinfo{person}{Murray Nguyen}, \bibinfo{person}{Murphy~Yuezhen Niu},
  \bibinfo{person}{Thomas~E. O'Brien}, \bibinfo{person}{Alex Opremcak}, \bibinfo{person}{John Platt}, \bibinfo{person}{Andre Petukhov}, \bibinfo{person}{Rebecca Potter}, \bibinfo{person}{Leonid~P. Pryadko}, \bibinfo{person}{Chris Quintana}, \bibinfo{person}{Pedram Roushan}, \bibinfo{person}{Nicholas~C. Rubin}, \bibinfo{person}{Negar Saei}, \bibinfo{person}{Daniel Sank}, \bibinfo{person}{Kannan Sankaragomathi}, \bibinfo{person}{Kevin~J. Satzinger}, \bibinfo{person}{Henry~F. Schurkus}, \bibinfo{person}{Christopher Schuster}, \bibinfo{person}{Michael~J. Shearn}, \bibinfo{person}{Aaron Shorter}, \bibinfo{person}{Vladimir Shvarts}, \bibinfo{person}{Jindra Skruzny}, \bibinfo{person}{Vadim Smelyanskiy}, \bibinfo{person}{W.~Clarke Smith}, \bibinfo{person}{George Sterling}, \bibinfo{person}{Doug Strain}, \bibinfo{person}{Marco Szalay}, \bibinfo{person}{Alfredo Torres}, \bibinfo{person}{Guifre Vidal}, \bibinfo{person}{Benjamin Villalonga}, \bibinfo{person}{Catherine Vollgraff~Heidweiller}, \bibinfo{person}{Theodore
  White}, \bibinfo{person}{Cheng Xing}, \bibinfo{person}{Z.~Jamie Yao}, \bibinfo{person}{Ping Yeh}, \bibinfo{person}{Juhwan Yoo}, \bibinfo{person}{Grayson Young}, \bibinfo{person}{Adam Zalcman}, \bibinfo{person}{Yaxing Zhang}, \bibinfo{person}{Ningfeng Zhu}, {and} \bibinfo{person}{Google~Quantum AI}.} \bibinfo{year}{2023}\natexlab{}.
\newblock \showarticletitle{Suppressing quantum errors by scaling a surface code logical qubit}.
\newblock \bibinfo{journal}{\emph{Nature}} \bibinfo{volume}{614}, \bibinfo{number}{7949} (\bibinfo{year}{2023}), \bibinfo{pages}{676--681}.
\newblock
\showISBNx{1476-4687}
\urldef\tempurl%
\url{https://doi.org/10.1038/s41586-022-05434-1}
\showDOI{\tempurl}


\bibitem[Arute et~al\mbox{.}(2019)]%
        {Google2019}
\bibfield{author}{\bibinfo{person}{Frank Arute}, \bibinfo{person}{Kunal Arya}, \bibinfo{person}{Ryan Babbush}, \bibinfo{person}{Dave Bacon}, \bibinfo{person}{Joseph~C. Bardin}, \bibinfo{person}{Rami Barends}, \bibinfo{person}{Rupak Biswas}, \bibinfo{person}{Sergio Boixo}, \bibinfo{person}{Fernando G. S.~L. Brandao}, \bibinfo{person}{David~A. Buell}, \bibinfo{person}{Brian Burkett}, \bibinfo{person}{Yu Chen}, \bibinfo{person}{Zijun Chen}, \bibinfo{person}{Ben Chiaro}, \bibinfo{person}{Roberto Collins}, \bibinfo{person}{William Courtney}, \bibinfo{person}{Andrew Dunsworth}, \bibinfo{person}{Edward Farhi}, \bibinfo{person}{Brooks Foxen}, \bibinfo{person}{Austin Fowler}, \bibinfo{person}{Craig Gidney}, \bibinfo{person}{Marissa Giustina}, \bibinfo{person}{Rob Graff}, \bibinfo{person}{Keith Guerin}, \bibinfo{person}{Steve Habegger}, \bibinfo{person}{Matthew~P. Harrigan}, \bibinfo{person}{Michael~J. Hartmann}, \bibinfo{person}{Alan Ho}, \bibinfo{person}{Markus Hoffmann}, \bibinfo{person}{Trent Huang},
  \bibinfo{person}{Travis~S. Humble}, \bibinfo{person}{Sergei~V. Isakov}, \bibinfo{person}{Evan Jeffrey}, \bibinfo{person}{Zhang Jiang}, \bibinfo{person}{Dvir Kafri}, \bibinfo{person}{Kostyantyn Kechedzhi}, \bibinfo{person}{Julian Kelly}, \bibinfo{person}{Paul~V. Klimov}, \bibinfo{person}{Sergey Knysh}, \bibinfo{person}{Alexander Korotkov}, \bibinfo{person}{Fedor Kostritsa}, \bibinfo{person}{David Landhuis}, \bibinfo{person}{Mike Lindmark}, \bibinfo{person}{Erik Lucero}, \bibinfo{person}{Dmitry Lyakh}, \bibinfo{person}{Salvatore Mandr{\`a}}, \bibinfo{person}{Jarrod~R. McClean}, \bibinfo{person}{Matthew McEwen}, \bibinfo{person}{Anthony Megrant}, \bibinfo{person}{Xiao Mi}, \bibinfo{person}{Kristel Michielsen}, \bibinfo{person}{Masoud Mohseni}, \bibinfo{person}{Josh Mutus}, \bibinfo{person}{Ofer Naaman}, \bibinfo{person}{Matthew Neeley}, \bibinfo{person}{Charles Neill}, \bibinfo{person}{Murphy~Yuezhen Niu}, \bibinfo{person}{Eric Ostby}, \bibinfo{person}{Andre Petukhov}, \bibinfo{person}{John~C. Platt},
  \bibinfo{person}{Chris Quintana}, \bibinfo{person}{Eleanor~G. Rieffel}, \bibinfo{person}{Pedram Roushan}, \bibinfo{person}{Nicholas~C. Rubin}, \bibinfo{person}{Daniel Sank}, \bibinfo{person}{Kevin~J. Satzinger}, \bibinfo{person}{Vadim Smelyanskiy}, \bibinfo{person}{Kevin~J. Sung}, \bibinfo{person}{Matthew~D. Trevithick}, \bibinfo{person}{Amit Vainsencher}, \bibinfo{person}{Benjamin Villalonga}, \bibinfo{person}{Theodore White}, \bibinfo{person}{Z.~Jamie Yao}, \bibinfo{person}{Ping Yeh}, \bibinfo{person}{Adam Zalcman}, \bibinfo{person}{Hartmut Neven}, {and} \bibinfo{person}{John~M. Martinis}.} \bibinfo{year}{2019}\natexlab{}.
\newblock \showarticletitle{Quantum supremacy using a programmable superconducting processor}.
\newblock \bibinfo{journal}{\emph{Nature}} \bibinfo{volume}{574}, \bibinfo{number}{7779} (\bibinfo{year}{2019}), \bibinfo{pages}{505--510}.
\newblock
\showISBNx{1476-4687}
\urldef\tempurl%
\url{https://doi.org/10.1038/s41586-019-1666-5}
\showDOI{\tempurl}


\bibitem[Bennink et~al\mbox{.}(2017)]%
        {bennink_unbiased_2017}
\bibfield{author}{\bibinfo{person}{Ryan~S. Bennink}, \bibinfo{person}{Erik~M. Ferragut}, \bibinfo{person}{Travis~S. Humble}, \bibinfo{person}{Jason~A. Laska}, \bibinfo{person}{James~J. Nutaro}, \bibinfo{person}{Mark~G. Pleszkoch}, {and} \bibinfo{person}{Raphael~C. Pooser}.} \bibinfo{year}{2017}\natexlab{}.
\newblock \showarticletitle{Unbiased {Simulation} of {Near}-{Clifford} {Quantum} {Circuits}}.
\newblock \bibinfo{journal}{\emph{Physical Review A}} \bibinfo{volume}{95}, \bibinfo{number}{6} (\bibinfo{date}{June} \bibinfo{year}{2017}), \bibinfo{pages}{062337}.
\newblock
\showISSN{2469-9926, 2469-9934}
\urldef\tempurl%
\url{https://doi.org/10.1103/PhysRevA.95.062337}
\showDOI{\tempurl}
\newblock
\shownote{arXiv:1703.00111 [quant-ph]}.


\bibitem[Bluvstein et~al\mbox{.}(2024)]%
        {Bluvstein2024}
\bibfield{author}{\bibinfo{person}{Dolev Bluvstein}, \bibinfo{person}{Simon~J. Evered}, \bibinfo{person}{Alexandra~A. Geim}, \bibinfo{person}{Sophie~H. Li}, \bibinfo{person}{Hengyun Zhou}, \bibinfo{person}{Tom Manovitz}, \bibinfo{person}{Sepehr Ebadi}, \bibinfo{person}{Madelyn Cain}, \bibinfo{person}{Marcin Kalinowski}, \bibinfo{person}{Dominik Hangleiter}, \bibinfo{person}{J.~Pablo Bonilla~Ataides}, \bibinfo{person}{Nishad Maskara}, \bibinfo{person}{Iris Cong}, \bibinfo{person}{Xun Gao}, \bibinfo{person}{Pedro Sales~Rodriguez}, \bibinfo{person}{Thomas Karolyshyn}, \bibinfo{person}{Giulia Semeghini}, \bibinfo{person}{Michael~J. Gullans}, \bibinfo{person}{Markus Greiner}, \bibinfo{person}{Vladan Vuleti{\'c}}, {and} \bibinfo{person}{Mikhail~D. Lukin}.} \bibinfo{year}{2024}\natexlab{}.
\newblock \showarticletitle{Logical quantum processor based on reconfigurable atom arrays}.
\newblock \bibinfo{journal}{\emph{Nature}} \bibinfo{volume}{626}, \bibinfo{number}{7997} (\bibinfo{year}{2024}), \bibinfo{pages}{58--65}.
\newblock
\showISBNx{1476-4687}
\urldef\tempurl%
\url{https://doi.org/10.1038/s41586-023-06927-3}
\showDOI{\tempurl}


\bibitem[Bravyi et~al\mbox{.}(2024a)]%
        {Bravyi2024}
\bibfield{author}{\bibinfo{person}{Sergey Bravyi}, \bibinfo{person}{Andrew~W. Cross}, \bibinfo{person}{Jay~M. Gambetta}, \bibinfo{person}{Dmitri Maslov}, \bibinfo{person}{Patrick Rall}, {and} \bibinfo{person}{Theodore~J. Yoder}.} \bibinfo{year}{2024}\natexlab{a}.
\newblock \showarticletitle{High-threshold and low-overhead fault-tolerant quantum memory}.
\newblock \bibinfo{journal}{\emph{Nature}} \bibinfo{volume}{627}, \bibinfo{number}{8005} (\bibinfo{year}{2024}), \bibinfo{pages}{778--782}.
\newblock
\showISBNx{1476-4687}
\urldef\tempurl%
\url{https://doi.org/10.1038/s41586-024-07107-7}
\showDOI{\tempurl}


\bibitem[Bravyi et~al\mbox{.}(2024b)]%
        {bravyi_high-threshold_2024}
\bibfield{author}{\bibinfo{person}{Sergey Bravyi}, \bibinfo{person}{Andrew~W. Cross}, \bibinfo{person}{Jay~M. Gambetta}, \bibinfo{person}{Dmitri Maslov}, \bibinfo{person}{Patrick Rall}, {and} \bibinfo{person}{Theodore~J. Yoder}.} \bibinfo{year}{2024}\natexlab{b}.
\newblock \showarticletitle{High-threshold and low-overhead fault-tolerant quantum memory}.
\newblock \bibinfo{journal}{\emph{Nature}} \bibinfo{volume}{627}, \bibinfo{number}{8005} (\bibinfo{date}{March} \bibinfo{year}{2024}), \bibinfo{pages}{778--782}.
\newblock
\showISSN{1476-4687}
\urldef\tempurl%
\url{https://doi.org/10.1038/s41586-024-07107-7}
\showDOI{\tempurl}
\newblock
\shownote{Publisher: Nature Publishing Group}.


\bibitem[Bravyi and Gosset(2016)]%
        {bravyi_improved_2016}
\bibfield{author}{\bibinfo{person}{Sergey Bravyi} {and} \bibinfo{person}{David Gosset}.} \bibinfo{year}{2016}\natexlab{}.
\newblock \showarticletitle{Improved {Classical} {Simulation} of {Quantum} {Circuits} {Dominated} by {Clifford} {Gates}}.
\newblock \bibinfo{journal}{\emph{Physical Review Letters}} \bibinfo{volume}{116}, \bibinfo{number}{25} (\bibinfo{date}{June} \bibinfo{year}{2016}), \bibinfo{pages}{250501}.
\newblock
\urldef\tempurl%
\url{https://doi.org/10.1103/PhysRevLett.116.250501}
\showDOI{\tempurl}
\newblock
\shownote{Publisher: American Physical Society}.


\bibitem[Bravyi et~al\mbox{.}(2016)]%
        {PBC}
\bibfield{author}{\bibinfo{person}{Sergey Bravyi}, \bibinfo{person}{Graeme Smith}, {and} \bibinfo{person}{John~A. Smolin}.} \bibinfo{year}{2016}\natexlab{}.
\newblock \showarticletitle{Trading Classical and Quantum Computational Resources}.
\newblock \bibinfo{journal}{\emph{Phys. Rev. X}}  \bibinfo{volume}{6} (\bibinfo{date}{Jun} \bibinfo{year}{2016}), \bibinfo{pages}{021043}.
\newblock
Issue 2.
\urldef\tempurl%
\url{https://doi.org/10.1103/PhysRevX.6.021043}
\showDOI{\tempurl}


\bibitem[Cerezo et~al\mbox{.}(2021)]%
        {Cerezo2021}
\bibfield{author}{\bibinfo{person}{M. Cerezo}, \bibinfo{person}{Andrew Arrasmith}, \bibinfo{person}{Ryan Babbush}, \bibinfo{person}{Simon~C. Benjamin}, \bibinfo{person}{Suguru Endo}, \bibinfo{person}{Keisuke Fujii}, \bibinfo{person}{Jarrod~R. McClean}, \bibinfo{person}{Kosuke Mitarai}, \bibinfo{person}{Xiao Yuan}, \bibinfo{person}{Lukasz Cincio}, {and} \bibinfo{person}{Patrick~J. Coles}.} \bibinfo{year}{2021}\natexlab{}.
\newblock \showarticletitle{Variational quantum algorithms}.
\newblock \bibinfo{journal}{\emph{Nature Reviews Physics}} \bibinfo{volume}{3}, \bibinfo{number}{9} (\bibinfo{year}{2021}), \bibinfo{pages}{625--644}.
\newblock
\showISBNx{2522-5820}
\urldef\tempurl%
\url{https://doi.org/10.1038/s42254-021-00348-9}
\showDOI{\tempurl}


\bibitem[Chen et~al\mbox{.}(2025)]%
        {chen_validating_2025}
\bibfield{author}{\bibinfo{person}{Kuan-Cheng Chen}, \bibinfo{person}{Tai-Yue Li}, \bibinfo{person}{Yun-Yuan Wang}, \bibinfo{person}{Simon See}, \bibinfo{person}{Chun-Chieh Wang}, \bibinfo{person}{Robert Wille}, \bibinfo{person}{Nan-Yow Chen}, \bibinfo{person}{An-Cheng Yang}, {and} \bibinfo{person}{Chun-Yu Lin}.} \bibinfo{year}{2025}\natexlab{}.
\newblock \bibinfo{title}{Validating {Large}-{Scale} {Quantum} {Machine} {Learning}: {Efficient} {Simulation} of {Quantum} {Support} {Vector} {Machines} {Using} {Tensor} {Networks}}.
\newblock
\newblock
\urldef\tempurl%
\url{https://doi.org/10.48550/arXiv.2405.02630}
\showDOI{\tempurl}
\newblock
\shownote{arXiv:2405.02630 [quant-ph] version: 3}.


\bibitem[Crawford et~al\mbox{.}(2021)]%
        {crawford_efficient_2021}
\bibfield{author}{\bibinfo{person}{Ophelia Crawford}, \bibinfo{person}{Barnaby~van Straaten}, \bibinfo{person}{Daochen Wang}, \bibinfo{person}{Thomas Parks}, \bibinfo{person}{Earl Campbell}, {and} \bibinfo{person}{Stephen Brierley}.} \bibinfo{year}{2021}\natexlab{}.
\newblock \bibinfo{title}{Efficient quantum measurement of {Pauli} operators in the presence of finite sampling error}.
\newblock
\newblock
\urldef\tempurl%
\url{https://doi.org/10.48550/arXiv.1908.06942}
\showDOI{\tempurl}
\newblock
\shownote{arXiv:1908.06942}.


\bibitem[Cross et~al\mbox{.}(2024)]%
        {cross2024improvedqldpcsurgerylogical}
\bibfield{author}{\bibinfo{person}{Andrew Cross}, \bibinfo{person}{Zhiyang He}, \bibinfo{person}{Patrick Rall}, {and} \bibinfo{person}{Theodore Yoder}.} \bibinfo{year}{2024}\natexlab{}.
\newblock \bibinfo{title}{Improved QLDPC Surgery: Logical Measurements and Bridging Codes}.
\newblock
\newblock
\showeprint[arxiv]{2407.18393}~[quant-ph]
\urldef\tempurl%
\url{https://arxiv.org/abs/2407.18393}
\showURL{%
\tempurl}


\bibitem[Dennis et~al\mbox{.}(2002)]%
        {Dennis2002}
\bibfield{author}{\bibinfo{person}{Eric Dennis}, \bibinfo{person}{Alexei Kitaev}, \bibinfo{person}{Andrew Landahl}, {and} \bibinfo{person}{John Preskill}.} \bibinfo{year}{2002}\natexlab{}.
\newblock \showarticletitle{Topological quantum memory}.
\newblock \bibinfo{journal}{\emph{J. Math. Phys.}} \bibinfo{volume}{43}, \bibinfo{number}{9} (\bibinfo{date}{09} \bibinfo{year}{2002}), \bibinfo{pages}{4452--4505}.
\newblock
\showISSN{0022-2488}
\urldef\tempurl%
\url{https://doi.org/10.1063/1.1499754}
\showDOI{\tempurl}
\showeprint{https://pubs.aip.org/aip/jmp/article-pdf/43/9/4452/19183135/4452\_1\_online.pdf}


\bibitem[Developers(2021)]%
        {cirq_developers_2021}
\bibfield{author}{\bibinfo{person}{Cirq Developers}.} \bibinfo{year}{2021}\natexlab{}.
\newblock \showarticletitle{Cirq: A Python Framework for Creating, Editing, and Invoking Noisy Intermediate Scale Quantum (NISQ) Circuits}.
\newblock \bibinfo{journal}{\emph{arXiv preprint arXiv:2008.08571}} (\bibinfo{year}{2021}).
\newblock
\urldef\tempurl%
\url{https://arxiv.org/abs/2008.08571}
\showURL{%
\tempurl}


\bibitem[Farhi et~al\mbox{.}(2014)]%
        {farhi2014quantum}
\bibfield{author}{\bibinfo{person}{Edward Farhi}, \bibinfo{person}{Jeffrey Goldstone}, {and} \bibinfo{person}{Sam Gutmann}.} \bibinfo{year}{2014}\natexlab{}.
\newblock \showarticletitle{A quantum approximate optimization algorithm}.
\newblock  (\bibinfo{year}{2014}).
\newblock
\urldef\tempurl%
\url{https://doi.org/10.48550/arXiv.1411.4028}
\showDOI{\tempurl}
\showeprint[arxiv]{arXiv:1411.4028}~[quant-ph]


\bibitem[Fowler et~al\mbox{.}(2012a)]%
        {Fowler2012}
\bibfield{author}{\bibinfo{person}{Austin~G. Fowler}, \bibinfo{person}{Matteo Mariantoni}, \bibinfo{person}{John~M. Martinis}, {and} \bibinfo{person}{Andrew~N. Cleland}.} \bibinfo{year}{2012}\natexlab{a}.
\newblock \showarticletitle{Surface codes: Towards practical large-scale quantum computation}.
\newblock \bibinfo{journal}{\emph{Phys. Rev. A}}  \bibinfo{volume}{86} (\bibinfo{date}{Sep} \bibinfo{year}{2012}), \bibinfo{pages}{032324}.
\newblock
Issue 3.
\urldef\tempurl%
\url{https://doi.org/10.1103/PhysRevA.86.032324}
\showDOI{\tempurl}


\bibitem[Fowler et~al\mbox{.}(2012b)]%
        {fowler_surface_2012}
\bibfield{author}{\bibinfo{person}{Austin~G. Fowler}, \bibinfo{person}{Matteo Mariantoni}, \bibinfo{person}{John~M. Martinis}, {and} \bibinfo{person}{Andrew~N. Cleland}.} \bibinfo{year}{2012}\natexlab{b}.
\newblock \showarticletitle{Surface codes: {Towards} practical large-scale quantum computation}.
\newblock \bibinfo{journal}{\emph{Physical Review A}} \bibinfo{volume}{86}, \bibinfo{number}{3} (\bibinfo{date}{Sept.} \bibinfo{year}{2012}), \bibinfo{pages}{032324}.
\newblock
\showISSN{1050-2947, 1094-1622}
\urldef\tempurl%
\url{https://doi.org/10.1103/PhysRevA.86.032324}
\showDOI{\tempurl}


\bibitem[Fowler et~al\mbox{.}(2012c)]%
        {Fowler2012_decoder_PRL}
\bibfield{author}{\bibinfo{person}{Austin~G. Fowler}, \bibinfo{person}{Adam~C. Whiteside}, {and} \bibinfo{person}{Lloyd C.~L. Hollenberg}.} \bibinfo{year}{2012}\natexlab{c}.
\newblock \showarticletitle{Towards Practical Classical Processing for the Surface Code}.
\newblock \bibinfo{journal}{\emph{Phys. Rev. Lett.}}  \bibinfo{volume}{108} (\bibinfo{date}{May} \bibinfo{year}{2012}), \bibinfo{pages}{180501}.
\newblock
Issue 18.
\urldef\tempurl%
\url{https://doi.org/10.1103/PhysRevLett.108.180501}
\showDOI{\tempurl}


\bibitem[Fowler et~al\mbox{.}(2012d)]%
        {Fowler2012_decoder_PRA}
\bibfield{author}{\bibinfo{person}{Austin~G. Fowler}, \bibinfo{person}{Adam~C. Whiteside}, {and} \bibinfo{person}{Lloyd C.~L. Hollenberg}.} \bibinfo{year}{2012}\natexlab{d}.
\newblock \showarticletitle{Towards practical classical processing for the surface code: Timing analysis}.
\newblock \bibinfo{journal}{\emph{Phys. Rev. A}}  \bibinfo{volume}{86} (\bibinfo{date}{Oct} \bibinfo{year}{2012}), \bibinfo{pages}{042313}.
\newblock
Issue 4.
\urldef\tempurl%
\url{https://doi.org/10.1103/PhysRevA.86.042313}
\showDOI{\tempurl}


\bibitem[Gao et~al\mbox{.}(2025)]%
        {Gao2025}
\bibfield{author}{\bibinfo{person}{Dongxin Gao}, \bibinfo{person}{Daojin Fan}, \bibinfo{person}{Chen Zha}, \bibinfo{person}{Jiahao Bei}, \bibinfo{person}{Guoqing Cai}, \bibinfo{person}{Jianbin Cai}, \bibinfo{person}{Sirui Cao}, \bibinfo{person}{Fusheng Chen}, \bibinfo{person}{Jiang Chen}, \bibinfo{person}{Kefu Chen}, \bibinfo{person}{Xiawei Chen}, \bibinfo{person}{Xiqing Chen}, \bibinfo{person}{Zhe Chen}, \bibinfo{person}{Zhiyuan Chen}, \bibinfo{person}{Zihua Chen}, \bibinfo{person}{Wenhao Chu}, \bibinfo{person}{Hui Deng}, \bibinfo{person}{Zhibin Deng}, \bibinfo{person}{Pei Ding}, \bibinfo{person}{Xun Ding}, \bibinfo{person}{Zhuzhengqi Ding}, \bibinfo{person}{Shuai Dong}, \bibinfo{person}{Yupeng Dong}, \bibinfo{person}{Bo Fan}, \bibinfo{person}{Yuanhao Fu}, \bibinfo{person}{Song Gao}, \bibinfo{person}{Lei Ge}, \bibinfo{person}{Ming Gong}, \bibinfo{person}{Jiacheng Gui}, \bibinfo{person}{Cheng Guo}, \bibinfo{person}{Shaojun Guo}, \bibinfo{person}{Xiaoyang Guo}, \bibinfo{person}{Lianchen Han},
  \bibinfo{person}{Tan He}, \bibinfo{person}{Linyin Hong}, \bibinfo{person}{Yisen Hu}, \bibinfo{person}{He-Liang Huang}, \bibinfo{person}{Yong-Heng Huo}, \bibinfo{person}{Tao Jiang}, \bibinfo{person}{Zuokai Jiang}, \bibinfo{person}{Honghong Jin}, \bibinfo{person}{Yunxiang Leng}, \bibinfo{person}{Dayu Li}, \bibinfo{person}{Dongdong Li}, \bibinfo{person}{Fangyu Li}, \bibinfo{person}{Jiaqi Li}, \bibinfo{person}{Jinjin Li}, \bibinfo{person}{Junyan Li}, \bibinfo{person}{Junyun Li}, \bibinfo{person}{Na Li}, \bibinfo{person}{Shaowei Li}, \bibinfo{person}{Wei Li}, \bibinfo{person}{Yuhuai Li}, \bibinfo{person}{Yuan Li}, \bibinfo{person}{Futian Liang}, \bibinfo{person}{Xuelian Liang}, \bibinfo{person}{Nanxing Liao}, \bibinfo{person}{Jin Lin}, \bibinfo{person}{Weiping Lin}, \bibinfo{person}{Dailin Liu}, \bibinfo{person}{Hongxiu Liu}, \bibinfo{person}{Maliang Liu}, \bibinfo{person}{Xinyu Liu}, \bibinfo{person}{Xuemeng Liu}, \bibinfo{person}{Yancheng Liu}, \bibinfo{person}{Haoxin Lou}, \bibinfo{person}{Yuwei Ma},
  \bibinfo{person}{Lingxin Meng}, \bibinfo{person}{Hao Mou}, \bibinfo{person}{Kailiang Nan}, \bibinfo{person}{Binghan Nie}, \bibinfo{person}{Meijuan Nie}, \bibinfo{person}{Jie Ning}, \bibinfo{person}{Le Niu}, \bibinfo{person}{Wenyi Peng}, \bibinfo{person}{Haoran Qian}, \bibinfo{person}{Hao Rong}, \bibinfo{person}{Tao Rong}, \bibinfo{person}{Huiyan Shen}, \bibinfo{person}{Qiong Shen}, \bibinfo{person}{Hong Su}, \bibinfo{person}{Feifan Su}, \bibinfo{person}{Chenyin Sun}, \bibinfo{person}{Liangchao Sun}, \bibinfo{person}{Tianzuo Sun}, \bibinfo{person}{Yingxiu Sun}, \bibinfo{person}{Yimeng Tan}, \bibinfo{person}{Jun Tan}, \bibinfo{person}{Longyue Tang}, \bibinfo{person}{Wenbing Tu}, \bibinfo{person}{Cai Wan}, \bibinfo{person}{Jiafei Wang}, \bibinfo{person}{Biao Wang}, \bibinfo{person}{Chang Wang}, \bibinfo{person}{Chen Wang}, \bibinfo{person}{Chu Wang}, \bibinfo{person}{Jian Wang}, \bibinfo{person}{Liangyuan Wang}, \bibinfo{person}{Rui Wang}, \bibinfo{person}{Shengtao Wang}, \bibinfo{person}{Xiaomin Wang},
  \bibinfo{person}{Xinzhe Wang}, \bibinfo{person}{Xunxun Wang}, \bibinfo{person}{Yeru Wang}, \bibinfo{person}{Zuolin Wei}, \bibinfo{person}{Jiazhou Wei}, \bibinfo{person}{Dachao Wu}, \bibinfo{person}{Gang Wu}, \bibinfo{person}{Jin Wu}, \bibinfo{person}{Shengjie Wu}, \bibinfo{person}{Yulin Wu}, \bibinfo{person}{Shiyong Xie}, \bibinfo{person}{Lianjie Xin}, \bibinfo{person}{Yu Xu}, \bibinfo{person}{Chun Xue}, \bibinfo{person}{Kai Yan}, \bibinfo{person}{Weifeng Yang}, \bibinfo{person}{Xinpeng Yang}, \bibinfo{person}{Yang Yang}, \bibinfo{person}{Yangsen Ye}, \bibinfo{person}{Zhenping Ye}, \bibinfo{person}{Chong Ying}, \bibinfo{person}{Jiale Yu}, \bibinfo{person}{Qinjing Yu}, \bibinfo{person}{Wenhu Yu}, \bibinfo{person}{Xiangdong Zeng}, \bibinfo{person}{Shaoyu Zhan}, \bibinfo{person}{Feifei Zhang}, \bibinfo{person}{Haibin Zhang}, \bibinfo{person}{Kaili Zhang}, \bibinfo{person}{Pan Zhang}, \bibinfo{person}{Wen Zhang}, \bibinfo{person}{Yiming Zhang}, \bibinfo{person}{Yongzhuo Zhang}, \bibinfo{person}{Lixiang Zhang},
  \bibinfo{person}{Guming Zhao}, \bibinfo{person}{Peng Zhao}, \bibinfo{person}{Xianhe Zhao}, \bibinfo{person}{Xintao Zhao}, \bibinfo{person}{Youwei Zhao}, \bibinfo{person}{Zhong Zhao}, \bibinfo{person}{Luyuan Zheng}, \bibinfo{person}{Fei Zhou}, \bibinfo{person}{Liang Zhou}, \bibinfo{person}{Na Zhou}, \bibinfo{person}{Naibin Zhou}, \bibinfo{person}{Shifeng Zhou}, \bibinfo{person}{Shuang Zhou}, \bibinfo{person}{Zhengxiao Zhou}, \bibinfo{person}{Chengjun Zhu}, \bibinfo{person}{Qingling Zhu}, \bibinfo{person}{Guihong Zou}, \bibinfo{person}{Haonan Zou}, \bibinfo{person}{Qiang Zhang}, \bibinfo{person}{Chao-Yang Lu}, \bibinfo{person}{Cheng-Zhi Peng}, \bibinfo{person}{Xiaobo Zhu}, {and} \bibinfo{person}{Jian-Wei Pan}.} \bibinfo{year}{2025}\natexlab{}.
\newblock \showarticletitle{Establishing a New Benchmark in Quantum Computational Advantage with 105-qubit Zuchongzhi 3.0 Processor}.
\newblock \bibinfo{journal}{\emph{Phys. Rev. Lett.}}  \bibinfo{volume}{134} (\bibinfo{date}{Mar} \bibinfo{year}{2025}), \bibinfo{pages}{090601}.
\newblock
Issue 9.
\urldef\tempurl%
\url{https://doi.org/10.1103/PhysRevLett.134.090601}
\showDOI{\tempurl}


\bibitem[Georgescu et~al\mbox{.}(2014)]%
        {georgescu2014quantum}
\bibfield{author}{\bibinfo{person}{Iulia~M Georgescu}, \bibinfo{person}{Sahel Ashhab}, {and} \bibinfo{person}{Franco Nori}.} \bibinfo{year}{2014}\natexlab{}.
\newblock \showarticletitle{Quantum simulation}.
\newblock \bibinfo{journal}{\emph{Reviews of Modern Physics}} \bibinfo{volume}{86}, \bibinfo{number}{1} (\bibinfo{year}{2014}), \bibinfo{pages}{153}.
\newblock


\bibitem[Gidney(2021)]%
        {gidney_stim_2021}
\bibfield{author}{\bibinfo{person}{Craig Gidney}.} \bibinfo{year}{2021}\natexlab{}.
\newblock \bibinfo{title}{Stim: a fast stabilizer circuit simulator}.
\newblock
\newblock
\urldef\tempurl%
\url{https://doi.org/10.48550/arXiv.2103.02202}
\showDOI{\tempurl}
\newblock
\shownote{arXiv:2103.02202}.


\bibitem[Gidney and Eker{\aa{}}(2021)]%
        {Gidney2021howtofactorbit}
\bibfield{author}{\bibinfo{person}{Craig Gidney} {and} \bibinfo{person}{Martin Eker{\aa{}}}.} \bibinfo{year}{2021}\natexlab{}.
\newblock \showarticletitle{How to factor 2048 bit {RSA} integers in 8 hours using 20 million noisy qubits}.
\newblock \bibinfo{journal}{\emph{{Quantum}}}  \bibinfo{volume}{5} (\bibinfo{date}{April} \bibinfo{year}{2021}), \bibinfo{pages}{433}.
\newblock
\showISSN{2521-327X}
\urldef\tempurl%
\url{https://doi.org/10.22331/q-2021-04-15-433}
\showDOI{\tempurl}


\bibitem[Gisin et~al\mbox{.}(2002)]%
        {gisin2002quantum}
\bibfield{author}{\bibinfo{person}{Nicolas Gisin}, \bibinfo{person}{Gr{\'e}goire Ribordy}, \bibinfo{person}{Wolfgang Tittel}, {and} \bibinfo{person}{Hugo Zbinden}.} \bibinfo{year}{2002}\natexlab{}.
\newblock \showarticletitle{Quantum cryptography}.
\newblock \bibinfo{journal}{\emph{Reviews of modern physics}} \bibinfo{volume}{74}, \bibinfo{number}{1} (\bibinfo{year}{2002}), \bibinfo{pages}{145}.
\newblock


\bibitem[Goto(2024)]%
        {Hayato2024}
\bibfield{author}{\bibinfo{person}{Hayato Goto}.} \bibinfo{year}{2024}\natexlab{}.
\newblock \showarticletitle{High-performance fault-tolerant quantum computing with many-hypercube codes}.
\newblock \bibinfo{journal}{\emph{Science Advances}} \bibinfo{volume}{10}, \bibinfo{number}{36} (\bibinfo{year}{2024}), \bibinfo{pages}{eadp6388}.
\newblock
\urldef\tempurl%
\url{https://doi.org/10.1126/sciadv.adp6388}
\showDOI{\tempurl}


\bibitem[Gottesman(1998)]%
        {gottesman_heisenberg_1998}
\bibfield{author}{\bibinfo{person}{Daniel Gottesman}.} \bibinfo{year}{1998}\natexlab{}.
\newblock \bibinfo{title}{The {Heisenberg} {Representation} of {Quantum} {Computers}}.
\newblock
\newblock
\urldef\tempurl%
\url{https://doi.org/10.48550/arXiv.quant-ph/9807006}
\showDOI{\tempurl}
\newblock
\shownote{arXiv:quant-ph/9807006}.


\bibitem[Han and Kim(2002)]%
        {han2002quantum}
\bibfield{author}{\bibinfo{person}{Kuk-Hyun Han} {and} \bibinfo{person}{Jong-Hwan Kim}.} \bibinfo{year}{2002}\natexlab{}.
\newblock \showarticletitle{Quantum-inspired evolutionary algorithm for a class of combinatorial optimization}.
\newblock \bibinfo{journal}{\emph{IEEE transactions on evolutionary computation}} \bibinfo{volume}{6}, \bibinfo{number}{6} (\bibinfo{year}{2002}), \bibinfo{pages}{580--593}.
\newblock


\bibitem[Higgott and Gidney(2025)]%
        {Higgott2025sparseblossom}
\bibfield{author}{\bibinfo{person}{Oscar Higgott} {and} \bibinfo{person}{Craig Gidney}.} \bibinfo{year}{2025}\natexlab{}.
\newblock \showarticletitle{Sparse {B}lossom: correcting a million errors per core second with minimum-weight matching}.
\newblock \bibinfo{journal}{\emph{{Quantum}}}  \bibinfo{volume}{9} (\bibinfo{date}{Jan.} \bibinfo{year}{2025}), \bibinfo{pages}{1600}.
\newblock
\showISSN{2521-327X}
\urldef\tempurl%
\url{https://doi.org/10.22331/q-2025-01-20-1600}
\showDOI{\tempurl}


\bibitem[Kandala et~al\mbox{.}(2017)]%
        {kandala2017hardware}
\bibfield{author}{\bibinfo{person}{Abhinav Kandala}, \bibinfo{person}{Antonio Mezzacapo}, \bibinfo{person}{Kristan Temme}, \bibinfo{person}{Maika Takita}, \bibinfo{person}{Markus Brink}, \bibinfo{person}{Jerry~M Chow}, {and} \bibinfo{person}{Jay~M Gambetta}.} \bibinfo{year}{2017}\natexlab{}.
\newblock \showarticletitle{Hardware-efficient variational quantum eigensolver for small molecules and quantum magnets}.
\newblock \bibinfo{journal}{\emph{Nature}} \bibinfo{volume}{549}, \bibinfo{number}{7671} (\bibinfo{year}{2017}), \bibinfo{pages}{242--246}.
\newblock


\bibitem[Katabarwa and Geller(2015)]%
        {katabarwa_logical_2015}
\bibfield{author}{\bibinfo{person}{Amara Katabarwa} {and} \bibinfo{person}{Michael~R. Geller}.} \bibinfo{year}{2015}\natexlab{}.
\newblock \showarticletitle{Logical error rate in the {Pauli} twirling approximation}.
\newblock \bibinfo{journal}{\emph{Scientific Reports}} \bibinfo{volume}{5}, \bibinfo{number}{1} (\bibinfo{date}{Sept.} \bibinfo{year}{2015}), \bibinfo{pages}{14670}.
\newblock
\showISSN{2045-2322}
\urldef\tempurl%
\url{https://doi.org/10.1038/srep14670}
\showDOI{\tempurl}


\bibitem[Katabarwa et~al\mbox{.}(2024)]%
        {Katabarwa2024}
\bibfield{author}{\bibinfo{person}{Amara Katabarwa}, \bibinfo{person}{Katerina Gratsea}, \bibinfo{person}{Athena Caesura}, {and} \bibinfo{person}{Peter~D. Johnson}.} \bibinfo{year}{2024}\natexlab{}.
\newblock \showarticletitle{Early Fault-Tolerant Quantum Computing}.
\newblock \bibinfo{journal}{\emph{PRX Quantum}}  \bibinfo{volume}{5} (\bibinfo{date}{Jun} \bibinfo{year}{2024}), \bibinfo{pages}{020101}.
\newblock
Issue 2.
\urldef\tempurl%
\url{https://doi.org/10.1103/PRXQuantum.5.020101}
\showDOI{\tempurl}


\bibitem[Kissinger and Wetering(2020)]%
        {kissinger_reducing_2020}
\bibfield{author}{\bibinfo{person}{Aleks Kissinger} {and} \bibinfo{person}{John van~de Wetering}.} \bibinfo{year}{2020}\natexlab{}.
\newblock \showarticletitle{Reducing {T}-count with the {ZX}-calculus}.
\newblock \bibinfo{journal}{\emph{Physical Review A}} \bibinfo{volume}{102}, \bibinfo{number}{2} (\bibinfo{date}{Aug.} \bibinfo{year}{2020}), \bibinfo{pages}{022406}.
\newblock
\showISSN{2469-9926, 2469-9934}
\urldef\tempurl%
\url{https://doi.org/10.1103/PhysRevA.102.022406}
\showDOI{\tempurl}
\newblock
\shownote{arXiv:1903.10477 [quant-ph]}.


\bibitem[Knill et~al\mbox{.}(2007)]%
        {knill_randomized_2007}
\bibfield{author}{\bibinfo{person}{E. Knill}, \bibinfo{person}{D. Leibfried}, \bibinfo{person}{R. Reichle}, \bibinfo{person}{J. Britton}, \bibinfo{person}{R.~B. Blakestad}, \bibinfo{person}{J.~D. Jost}, \bibinfo{person}{C. Langer}, \bibinfo{person}{R. Ozeri}, \bibinfo{person}{S. Seidelin}, {and} \bibinfo{person}{D.~J. Wineland}.} \bibinfo{year}{2007}\natexlab{}.
\newblock \bibinfo{title}{Randomized {Benchmarking} of {Quantum} {Gates}}.
\newblock
\newblock
\urldef\tempurl%
\url{https://doi.org/10.1103/PhysRevA.77.012307}
\showDOI{\tempurl}


\bibitem[Krinner et~al\mbox{.}(2022)]%
        {Krinner2022}
\bibfield{author}{\bibinfo{person}{Sebastian Krinner}, \bibinfo{person}{Nathan Lacroix}, \bibinfo{person}{Ants Remm}, \bibinfo{person}{Agustin Di~Paolo}, \bibinfo{person}{Elie Genois}, \bibinfo{person}{Catherine Leroux}, \bibinfo{person}{Christoph Hellings}, \bibinfo{person}{Stefania Lazar}, \bibinfo{person}{Francois Swiadek}, \bibinfo{person}{Johannes Herrmann}, \bibinfo{person}{Graham~J. Norris}, \bibinfo{person}{Christian~Kraglund Andersen}, \bibinfo{person}{Markus M{\"u}ller}, \bibinfo{person}{Alexandre Blais}, \bibinfo{person}{Christopher Eichler}, {and} \bibinfo{person}{Andreas Wallraff}.} \bibinfo{year}{2022}\natexlab{}.
\newblock \showarticletitle{Realizing repeated quantum error correction in a distance-three surface code}.
\newblock \bibinfo{journal}{\emph{Nature}} \bibinfo{volume}{605}, \bibinfo{number}{7911} (\bibinfo{year}{2022}), \bibinfo{pages}{669--674}.
\newblock
\showISBNx{1476-4687}
\urldef\tempurl%
\url{https://doi.org/10.1038/s41586-022-04566-8}
\showDOI{\tempurl}


\bibitem[Litinski(2019)]%
        {litinski_game_2019}
\bibfield{author}{\bibinfo{person}{Daniel Litinski}.} \bibinfo{year}{2019}\natexlab{}.
\newblock \showarticletitle{A {Game} of {Surface} {Codes}: {Large}-{Scale} {Quantum} {Computing} with {Lattice} {Surgery}}.
\newblock \bibinfo{journal}{\emph{Quantum}}  \bibinfo{volume}{3} (\bibinfo{date}{March} \bibinfo{year}{2019}), \bibinfo{pages}{128}.
\newblock
\showISSN{2521-327X}
\urldef\tempurl%
\url{https://doi.org/10.22331/q-2019-03-05-128}
\showDOI{\tempurl}
\newblock
\shownote{arXiv:1808.02892 [quant-ph]}.


\bibitem[Nielsen and Chuang(2010)]%
        {nielsen_chuang_2010}
\bibfield{author}{\bibinfo{person}{Michael~A. Nielsen} {and} \bibinfo{person}{Isaac~L. Chuang}.} \bibinfo{year}{2010}\natexlab{}.
\newblock \bibinfo{booktitle}{\emph{Quantum Computation and Quantum Information: 10th Anniversary Edition}}.
\newblock \bibinfo{publisher}{Cambridge University Press}.
\newblock
\urldef\tempurl%
\url{https://doi.org/10.1017/CBO9780511976667}
\showDOI{\tempurl}


\bibitem[{NVIDIA Corporation \& Affiliates}({[n.\,d.]})]%
        {cudaqx}
\bibfield{author}{\bibinfo{person}{{NVIDIA Corporation \& Affiliates}}.} \bibinfo{year}{[n.\,d.]}\natexlab{}.
\newblock \bibinfo{booktitle}{\emph{CUDA‑QX}}.
\newblock
\urldef\tempurl%
\url{https://github.com/NVIDIA/cudaqx}
\showURL{%
\tempurl}
\newblock
\shownote{If you use CUDA‑QX in your work, please also cite CUDA‑Q (https://github.com/NVIDIA/cuda-quantum).}.


\bibitem[Osama et~al\mbox{.}(2025)]%
        {osama_parallel_2025}
\bibfield{author}{\bibinfo{person}{Muhammad Osama}, \bibinfo{person}{Dimitrios Thanos}, {and} \bibinfo{person}{Alfons Laarman}.} \bibinfo{year}{2025}\natexlab{}.
\newblock \showarticletitle{Parallel {Equivalence} {Checking} of {Stabilizer} {Quantum} {Circuits} on {GPUs}}. In \bibinfo{booktitle}{\emph{Tools and {Algorithms} for the {Construction} and {Analysis} of {Systems}}}, \bibfield{editor}{\bibinfo{person}{Arie Gurfinkel} {and} \bibinfo{person}{Marijn Heule}} (Eds.). \bibinfo{publisher}{Springer Nature Switzerland}, \bibinfo{address}{Cham}, \bibinfo{pages}{109--128}.
\newblock
\showISBNx{9783031906602}
\urldef\tempurl%
\url{https://doi.org/10.1007/978-3-031-90660-2_6}
\showDOI{\tempurl}


\bibitem[Peres and Galvão(2023)]%
        {peres_quantum_2023}
\bibfield{author}{\bibinfo{person}{Filipa C.~R. Peres} {and} \bibinfo{person}{Ernesto~F. Galvão}.} \bibinfo{year}{2023}\natexlab{}.
\newblock \showarticletitle{Quantum circuit compilation and hybrid computation using {Pauli}-based computation}.
\newblock \bibinfo{journal}{\emph{Quantum}}  \bibinfo{volume}{7} (\bibinfo{date}{Oct.} \bibinfo{year}{2023}), \bibinfo{pages}{1126}.
\newblock
\showISSN{2521-327X}
\urldef\tempurl%
\url{https://doi.org/10.22331/q-2023-10-03-1126}
\showDOI{\tempurl}
\newblock
\shownote{arXiv:2203.01789 [quant-ph]}.


\bibitem[Peruzzo et~al\mbox{.}(2014)]%
        {Peruzzo2014}
\bibfield{author}{\bibinfo{person}{Alberto Peruzzo}, \bibinfo{person}{Jarrod McClean}, \bibinfo{person}{Peter Shadbolt}, \bibinfo{person}{Man-Hong Yung}, \bibinfo{person}{Xiao-Qi Zhou}, \bibinfo{person}{Peter~J. Love}, \bibinfo{person}{Al{\'a}n Aspuru-Guzik}, {and} \bibinfo{person}{Jeremy~L. O'Brien}.} \bibinfo{year}{2014}\natexlab{}.
\newblock \showarticletitle{A variational eigenvalue solver on a photonic quantum processor}.
\newblock \bibinfo{journal}{\emph{Nature Communications}} \bibinfo{volume}{5}, \bibinfo{number}{1} (\bibinfo{year}{2014}), \bibinfo{pages}{4213}.
\newblock
\showISBNx{2041-1723}
\urldef\tempurl%
\url{https://doi.org/10.1038/ncomms5213}
\showDOI{\tempurl}


\bibitem[Preskill(2018)]%
        {NISQ_Preskill}
\bibfield{author}{\bibinfo{person}{John Preskill}.} \bibinfo{year}{2018}\natexlab{}.
\newblock \showarticletitle{Quantum {C}omputing in the {NISQ} era and beyond}.
\newblock \bibinfo{journal}{\emph{{Quantum}}}  \bibinfo{volume}{2} (\bibinfo{date}{Aug.} \bibinfo{year}{2018}), \bibinfo{pages}{79}.
\newblock
\showISSN{2521-327X}
\urldef\tempurl%
\url{https://doi.org/10.22331/q-2018-08-06-79}
\showDOI{\tempurl}


\bibitem[Preskill(2025)]%
        {Preskill_megaquop}
\bibfield{author}{\bibinfo{person}{John Preskill}.} \bibinfo{year}{2025}\natexlab{}.
\newblock \showarticletitle{Beyond NISQ: The Megaquop Machine}.
\newblock \bibinfo{journal}{\emph{ACM Transactions on Quantum Computing}} (\bibinfo{date}{March} \bibinfo{year}{2025}).
\newblock
\urldef\tempurl%
\url{https://doi.org/10.1145/3723153}
\showDOI{\tempurl}
\newblock
\shownote{Just Accepted}.


\bibitem[{Qiskit Development Team}(2021)]%
        {qiskit}
\bibfield{author}{\bibinfo{person}{{Qiskit Development Team}}.} \bibinfo{year}{2021}\natexlab{}.
\newblock \showarticletitle{{Qiskit: An Open-source Framework for Quantum Computing}}.
\newblock \bibinfo{journal}{\emph{Zenodo}} (\bibinfo{year}{2021}).
\newblock
\urldef\tempurl%
\url{https://doi.org/10.5281/zenodo.2562111}
\showDOI{\tempurl}


\bibitem[Ryan-Anderson et~al\mbox{.}(2021)]%
        {Anderson2021}
\bibfield{author}{\bibinfo{person}{C. Ryan-Anderson}, \bibinfo{person}{J.~G. Bohnet}, \bibinfo{person}{K. Lee}, \bibinfo{person}{D. Gresh}, \bibinfo{person}{A. Hankin}, \bibinfo{person}{J.~P. Gaebler}, \bibinfo{person}{D. Francois}, \bibinfo{person}{A. Chernoguzov}, \bibinfo{person}{D. Lucchetti}, \bibinfo{person}{N.~C. Brown}, \bibinfo{person}{T.~M. Gatterman}, \bibinfo{person}{S.~K. Halit}, \bibinfo{person}{K. Gilmore}, \bibinfo{person}{J.~A. Gerber}, \bibinfo{person}{B. Neyenhuis}, \bibinfo{person}{D. Hayes}, {and} \bibinfo{person}{R.~P. Stutz}.} \bibinfo{year}{2021}\natexlab{}.
\newblock \showarticletitle{Realization of Real-Time Fault-Tolerant Quantum Error Correction}.
\newblock \bibinfo{journal}{\emph{Phys. Rev. X}}  \bibinfo{volume}{11} (\bibinfo{date}{Dec} \bibinfo{year}{2021}), \bibinfo{pages}{041058}.
\newblock
Issue 4.
\urldef\tempurl%
\url{https://doi.org/10.1103/PhysRevX.11.041058}
\showDOI{\tempurl}


\bibitem[Scriva et~al\mbox{.}(2024)]%
        {scriva_challenges_2024}
\bibfield{author}{\bibinfo{person}{Giuseppe Scriva}, \bibinfo{person}{Nikita Astrakhantsev}, \bibinfo{person}{Sebastiano Pilati}, {and} \bibinfo{person}{Guglielmo Mazzola}.} \bibinfo{year}{2024}\natexlab{}.
\newblock \showarticletitle{Challenges of variational quantum optimization with measurement shot noise}.
\newblock \bibinfo{journal}{\emph{Physical Review A}} \bibinfo{volume}{109}, \bibinfo{number}{3} (\bibinfo{date}{March} \bibinfo{year}{2024}), \bibinfo{pages}{032408}.
\newblock
\showISSN{2469-9926, 2469-9934}
\urldef\tempurl%
\url{https://doi.org/10.1103/PhysRevA.109.032408}
\showDOI{\tempurl}
\newblock
\shownote{arXiv:2308.00044 [quant-ph]}.


\bibitem[Shor(1999)]%
        {shor1999polynomial}
\bibfield{author}{\bibinfo{person}{Peter~W Shor}.} \bibinfo{year}{1999}\natexlab{}.
\newblock \showarticletitle{Polynomial-time algorithms for prime factorization and discrete logarithms on a quantum computer}.
\newblock \bibinfo{journal}{\emph{SIAM review}} \bibinfo{volume}{41}, \bibinfo{number}{2} (\bibinfo{year}{1999}), \bibinfo{pages}{303--332}.
\newblock


\bibitem[Stein et~al\mbox{.}(2024)]%
        {stein_architectures_2024}
\bibfield{author}{\bibinfo{person}{Samuel Stein}, \bibinfo{person}{Shifan Xu}, \bibinfo{person}{Andrew~W. Cross}, \bibinfo{person}{Theodore~J. Yoder}, \bibinfo{person}{Ali Javadi-Abhari}, \bibinfo{person}{Chenxu Liu}, \bibinfo{person}{Kun Liu}, \bibinfo{person}{Zeyuan Zhou}, \bibinfo{person}{Charles Guinn}, \bibinfo{person}{Yufei Ding}, \bibinfo{person}{Yongshan Ding}, {and} \bibinfo{person}{Ang Li}.} \bibinfo{year}{2024}\natexlab{}.
\newblock \bibinfo{title}{Architectures for {Heterogeneous} {Quantum} {Error} {Correction} {Codes}}.
\newblock
\newblock
\urldef\tempurl%
\url{https://doi.org/10.48550/arXiv.2411.03202}
\showDOI{\tempurl}
\newblock
\shownote{arXiv:2411.03202 [quant-ph]}.


\bibitem[Tilly et~al\mbox{.}(2022)]%
        {tilly_variational_2022}
\bibfield{author}{\bibinfo{person}{Jules Tilly}, \bibinfo{person}{Hongxiang Chen}, \bibinfo{person}{Shuxiang Cao}, \bibinfo{person}{Dario Picozzi}, \bibinfo{person}{Kanav Setia}, \bibinfo{person}{Ying Li}, \bibinfo{person}{Edward Grant}, \bibinfo{person}{Leonard Wossnig}, \bibinfo{person}{Ivan Rungger}, \bibinfo{person}{George~H. Booth}, {and} \bibinfo{person}{Jonathan Tennyson}.} \bibinfo{year}{2022}\natexlab{}.
\newblock \showarticletitle{The {Variational} {Quantum} {Eigensolver}: a review of methods and best practices}.
\newblock \bibinfo{journal}{\emph{Physics Reports}}  \bibinfo{volume}{986} (\bibinfo{date}{Nov.} \bibinfo{year}{2022}), \bibinfo{pages}{1--128}.
\newblock
\showISSN{03701573}
\urldef\tempurl%
\url{https://doi.org/10.1016/j.physrep.2022.08.003}
\showDOI{\tempurl}
\newblock
\shownote{arXiv:2111.05176 [quant-ph]}.


\bibitem[Tomita and Svore(2014)]%
        {tomita_low-distance_2014}
\bibfield{author}{\bibinfo{person}{Yu Tomita} {and} \bibinfo{person}{Krysta~M. Svore}.} \bibinfo{year}{2014}\natexlab{}.
\newblock \showarticletitle{Low-distance surface codes under realistic quantum noise}.
\newblock \bibinfo{journal}{\emph{Physical Review A}} \bibinfo{volume}{90}, \bibinfo{number}{6} (\bibinfo{date}{Dec.} \bibinfo{year}{2014}), \bibinfo{pages}{062320}.
\newblock
\urldef\tempurl%
\url{https://doi.org/10.1103/PhysRevA.90.062320}
\showDOI{\tempurl}


\bibitem[Verteletskyi et~al\mbox{.}(2020)]%
        {verteletskyi_measurement_2020}
\bibfield{author}{\bibinfo{person}{Vladyslav Verteletskyi}, \bibinfo{person}{Tzu-Ching Yen}, {and} \bibinfo{person}{Artur~F. Izmaylov}.} \bibinfo{year}{2020}\natexlab{}.
\newblock \showarticletitle{Measurement {Optimization} in the {Variational} {Quantum} {Eigensolver} {Using} a {Minimum} {Clique} {Cover}}.
\newblock \bibinfo{journal}{\emph{The Journal of Chemical Physics}} \bibinfo{volume}{152}, \bibinfo{number}{12} (\bibinfo{date}{March} \bibinfo{year}{2020}), \bibinfo{pages}{124114}.
\newblock
\showISSN{0021-9606, 1089-7690}
\urldef\tempurl%
\url{https://doi.org/10.1063/1.5141458}
\showDOI{\tempurl}
\newblock
\shownote{arXiv:1907.03358 [quant-ph]}.


\bibitem[Wetering(2020)]%
        {wetering_zx-calculus_2020}
\bibfield{author}{\bibinfo{person}{John van~de Wetering}.} \bibinfo{year}{2020}\natexlab{}.
\newblock \bibinfo{title}{{ZX}-calculus for the working quantum computer scientist}.
\newblock
\newblock
\urldef\tempurl%
\url{https://doi.org/10.48550/arXiv.2012.13966}
\showDOI{\tempurl}
\newblock
\shownote{arXiv:2012.13966 [quant-ph]}.


\bibitem[Wu et~al\mbox{.}(2021)]%
        {Pan2021}
\bibfield{author}{\bibinfo{person}{Yulin Wu}, \bibinfo{person}{Wan-Su Bao}, \bibinfo{person}{Sirui Cao}, \bibinfo{person}{Fusheng Chen}, \bibinfo{person}{Ming-Cheng Chen}, \bibinfo{person}{Xiawei Chen}, \bibinfo{person}{Tung-Hsun Chung}, \bibinfo{person}{Hui Deng}, \bibinfo{person}{Yajie Du}, \bibinfo{person}{Daojin Fan}, \bibinfo{person}{Ming Gong}, \bibinfo{person}{Cheng Guo}, \bibinfo{person}{Chu Guo}, \bibinfo{person}{Shaojun Guo}, \bibinfo{person}{Lianchen Han}, \bibinfo{person}{Linyin Hong}, \bibinfo{person}{He-Liang Huang}, \bibinfo{person}{Yong-Heng Huo}, \bibinfo{person}{Liping Li}, \bibinfo{person}{Na Li}, \bibinfo{person}{Shaowei Li}, \bibinfo{person}{Yuan Li}, \bibinfo{person}{Futian Liang}, \bibinfo{person}{Chun Lin}, \bibinfo{person}{Jin Lin}, \bibinfo{person}{Haoran Qian}, \bibinfo{person}{Dan Qiao}, \bibinfo{person}{Hao Rong}, \bibinfo{person}{Hong Su}, \bibinfo{person}{Lihua Sun}, \bibinfo{person}{Liangyuan Wang}, \bibinfo{person}{Shiyu Wang}, \bibinfo{person}{Dachao Wu}, \bibinfo{person}{Yu
  Xu}, \bibinfo{person}{Kai Yan}, \bibinfo{person}{Weifeng Yang}, \bibinfo{person}{Yang Yang}, \bibinfo{person}{Yangsen Ye}, \bibinfo{person}{Jianghan Yin}, \bibinfo{person}{Chong Ying}, \bibinfo{person}{Jiale Yu}, \bibinfo{person}{Chen Zha}, \bibinfo{person}{Cha Zhang}, \bibinfo{person}{Haibin Zhang}, \bibinfo{person}{Kaili Zhang}, \bibinfo{person}{Yiming Zhang}, \bibinfo{person}{Han Zhao}, \bibinfo{person}{Youwei Zhao}, \bibinfo{person}{Liang Zhou}, \bibinfo{person}{Qingling Zhu}, \bibinfo{person}{Chao-Yang Lu}, \bibinfo{person}{Cheng-Zhi Peng}, \bibinfo{person}{Xiaobo Zhu}, {and} \bibinfo{person}{Jian-Wei Pan}.} \bibinfo{year}{2021}\natexlab{}.
\newblock \showarticletitle{Strong Quantum Computational Advantage Using a Superconducting Quantum Processor}.
\newblock \bibinfo{journal}{\emph{Phys. Rev. Lett.}}  \bibinfo{volume}{127} (\bibinfo{date}{Oct} \bibinfo{year}{2021}), \bibinfo{pages}{180501}.
\newblock
Issue 18.
\urldef\tempurl%
\url{https://doi.org/10.1103/PhysRevLett.127.180501}
\showDOI{\tempurl}


\bibitem[Yen et~al\mbox{.}(2023)]%
        {Yen2023}
\bibfield{author}{\bibinfo{person}{Tzu-Ching Yen}, \bibinfo{person}{Aadithya Ganeshram}, {and} \bibinfo{person}{Artur~F. Izmaylov}.} \bibinfo{year}{2023}\natexlab{}.
\newblock \showarticletitle{Deterministic improvements of quantum measurements with grouping of compatible operators, non-local transformations, and covariance estimates}.
\newblock \bibinfo{journal}{\emph{npj Quantum Information}} \bibinfo{volume}{9}, \bibinfo{number}{1} (\bibinfo{year}{2023}), \bibinfo{pages}{14}.
\newblock
\showISBNx{2056-6387}
\urldef\tempurl%
\url{https://doi.org/10.1038/s41534-023-00683-y}
\showDOI{\tempurl}


\end{thebibliography}
\appendix
\section{Appendix}

This appendix provides detailed pseudocode for the algorithms used in transpilation, the T-optimization data set, and the original CHP $rowsum$ algorithm for convenience to refer to throughout.

\subsection{Tableau T-Transpiler Algorithms}
\label{app:t_algs}
The following are the three key stages described in Section~\ref{sec:push_design} and transcribed here into pseudocode for further reading.

\begin{algorithm} [h]
\caption{Tableau Construction}
\label{alg:t_con}
\begin{algorithmic}[1]
\State Initialize an empty tableau $T_{tab}$ to store $pi/4$ rotation gates ($T$ and $T^\dagger$), and an $n \times n$ identity tableau $M_{tab}$.
\ForAll{gates $G$ at qubit $q$ starting from the end}
    \If{$G = T_q$}
        \State Append an $n$-qubit $Z_{q}$-stabilizer to $T_{tab}$ and set $r \gets 0$
    \ElsIf{$G = T^\dagger_q$}
        \State Append an $n$-qubit $Z_{q}$-stabilizer to $T_{tab}$ and set $r \gets 1$
    \ElsIf{$G$ is Clifford}
        \State Apply $G$ to $M_{tab}$ and $T_{tab}$ using CHP rules
    \EndIf
\EndFor
\end{algorithmic}
\end{algorithm}

\begin{algorithm} [h]
\caption{T-Separation}
\label{alg:t_sep}
\begin{algorithmic}[1]
\State Initialize tableau list $\mathcal{P} \gets [\ ]$
\State Create tableau $P_0$ and append the last stabilizer in $T_{\text{tab}}$
\State Append $P_0$ to $\mathcal{P}$

\ForAll{stabilizers $S$ from second-to-last to first in $T_{\text{tab}}$}
    \State $appended \gets \textbf{false}$
    \ForAll{$P \in \mathcal{P}$ starting from $P_{0}$}
        \If{$[S, P] = 0$}
            \State Append $S$ to $P$
            \State $appended \gets \textbf{true}$
            \State \textbf{break}
        \EndIf
    \EndFor
    \If{\textbf{not} $appended$}
        \State Create new tableau $P_{\lvert \mathcal{P} \rvert}$ and append $S$
        \State Append $P_{\lvert \mathcal{P} \rvert}$ to $\mathcal{P}$
    \EndIf
\EndFor
\end{algorithmic}
\end{algorithm}

\begin{algorithm}
\caption{T-Optimization}
\label{alg:t_opt}
\begin{algorithmic}
\State Count initial total number of rows in all $P_i \in \mathcal{P}$, store as $N_{\text{prev}}$
\State Initialize convergence flag $converged \gets \textbf{false}$

\While{\textbf{not} $converged$}
    \For{$i = \lvert \mathcal{P} \rvert - 1$ down to $0$}
        \While{there exist two identical stabilizers $S$ in $P_i$}
            \State Remove both copies of $S$ from $P_i$
            \State Append one copy of $S$ to $P_{i+1}$
            \ForAll{stabilizers $S_{T}$ in $P_{i+1}$}
                \If{$[S_{T}, S] \neq 0$}
                    \State Apply $\text{rowsum}(T, S)$
                \EndIf
            \EndFor
            \State Remove $S$ from $P_{i+1}$ and append to $P_{i+2}$
        \EndWhile
    \EndFor
    
    \State Count the total number of rows in all $P_i \in \mathcal{P}$, store as $N_{\text{new}}$
    
    \If{$N_{\text{new}} = N_{\text{prev}}$}
        \State $converged \gets \textbf{true}$
    \Else
        \State $N_{\text{prev}} \gets N_{\text{new}}$
    \EndIf
\EndWhile
\State $\lvert \mathcal{P} \rvert$ is the number of $\pi/8$ rotation gate layers after minimization. The sum of all non-Identity Pauli operators in all $P_{i}$ stabilizers is the number of $\pi/8$ rotation gates.
\end{algorithmic}
\end{algorithm}

\newpage
\subsection{Rowsum Algorithm}
\label{app:rowsum}

The original CHP $rowsum$ algorithm is key to understanding the parallel implementation used to determine stabilizer phase. This algorithm makes use of a helper function $g$ for computing local commutation contributions. 

\begin{algorithm}[h]
\caption{\textsc{rowsum}$(h, i)$}
\label{alg:rowsum}
\begin{algorithmic}[1]
\Function{g}{$x_1, z_1, x_2, z_2$}
    \If{$x_1 = 0$ and $z_1 = 0$}
        \State \Return $0$
    \ElsIf{$x_1 = 1$ and $z_1 = 1$}
        \State \Return $z_2 - x_2$
    \ElsIf{$x_1 = 1$ and $z_1 = 0$}
        \State \Return $z_2 \cdot (2x_2 - 1)$
    \ElsIf{$x_1 = 0$ and $z_1 = 1$}
        \State \Return $x_2 \cdot (1 - 2z_2)$
    \EndIf
\EndFunction

\Statex

\Function{rowsum}{$h, i$}
    \State $sum \gets 2r_h + 2r_i$
    \For{$j = 1$ to $n$}
        \State $sum \gets sum + \Call{g}{x_{ij}, z_{ij}, x_{hj}, z_{hj}}$
    \EndFor
    \If{$sum \bmod 4 = 0$}
        \State $r_h \gets 0$
    \ElsIf{$sum \bmod 4 = 2$}
        \State $r_h \gets 1$
    \EndIf
    \For{$j = 1$ to $n$}
        \State $x_{hj} \gets x_{ij} \oplus x_{hj}$
        \State $z_{hj} \gets z_{ij} \oplus z_{hj}$
    \EndFor
\EndFunction
\end{algorithmic}
\end{algorithm}

\newpage
\subsection{Circuit Optimization Results}
\label{app:opt_results}

Here we include extended evaluation data for Clifford+T benchmark circuits.  
Table~\ref{tab:circuit-perf} lists the detailed Clifford+T gate counts and relative speedup of \framework versus our own Python reference, both using Algorithms~\ref{alg:t_con},~\ref{alg:t_sep}, and~\ref{alg:t_opt}. A more concise comparison to PyZX is provided in Figure~\ref{fig:t_comp}.

\begin{table}[H]
\centering
\scriptsize
\setlength{\tabcolsep}{4pt} 
\begin{tabular}{lrrrr}
\toprule
Circuit & Python Time & Total Gates & T-Gates: Opt-T & T-Reduction \\
\midrule
Adder N10 & 14.26x & 189 & 88:16 & 5.5x \\
Adder N4 & 34.61x & 25 & 4:4 & 1.0x \\
Adder N433 & 2.74x & 6196 & 1536:768 & 2.0x \\
Basis Test N4 & 51.88x & 577 & 156:132 & 1.18x \\
Basis Trotter N4 & 16.64x & 9819 & 1860:1578 & 1.18x \\
Bell N4 & 17.19x & 386 & 70:68 & 1.03x \\
Dnn N16 & 8.17x & 30016 & 6656:5920 & 1.12x \\
Dnn N2 & 46.46x & 3414 & 760:672 & 1.13x \\
Dnn N8 & 16.56x & 15008 & 3328:2960 & 1.12x \\
Fredkin N3 & 19.28x & 22 & 4:4 & 1.0x \\
Gcm H6 & 7.97x & 17719 & 4852:4486 & 1.08x \\
Inverseqft N4 & 13.29x & 99 & 28:28 & 1.0x \\
Ising N10 & 30.14x & 3152 & 974:928 & 1.05x \\
Ising N26 & 43.93x & 1116 & 332:324 & 1.02x \\
Knn N25 & 11.79x & 816 & 204:192 & 1.06x \\
Multiplier N15 & 3.32x & 454 & 144:42 & 3.43x \\
Multiplier N45 & 6.77x & 4574 & 1512:396 & 3.82x \\
Multiply N13 & 11.67x & 92 & 24:14 & 1.71x \\
Pea N5 & 13.69x & 523 & 166:166 & 1.0x \\
Qaoa N3 & 28.38x & 53 & 8:8 & 1.0x \\
Qaoa N6 & 28.65x & 2928 & 672:544 & 1.24x \\
Qec En N5 & 27.47x & 25 & 1:1 & 1.0x \\
Qf21 N15 & 18.42x & 646 & 170:162 & 1.05x \\
Qft N18 & 19.05x & 2150 & 596:596 & 1.0x \\
Qft N18 Iter1 & 29.37x & 3290 & 1677:1333 & 1.26x \\
Qft N18 Iter2 & 20.16x & 27292 & 14383:11463 & 1.25x \\
Qft N4 & 38.83x & 310 & 92:92 & 1.0x \\
Qpe N9 & 53.14x & 334 & 90:86 & 1.05x \\
Qram N20 & 32.09x & 276 & 80:48 & 1.67x \\
Sat N7 & 32.62x & 213 & 40:24 & 1.67x \\
Seca N11 & 29.27x & 200 & 32:16 & 2.0x \\
Simon N6 & 21.12x & 56 & 8:8 & 1.0x \\
Swap Test N115 & 10.99x & 2866 & 736:680 & 1.08x \\
Swap Test N25 & 13.76x & 592 & 152:140 & 1.09x \\
Teleportation N3 & 56.61x & 8 & 1:1 & 1.0x \\
Toffoli N3 & 22.36x & 20 & 3:3 & 1.0x \\
Variational N4 & 51.23x & 363 & 100:92 & 1.09x \\
Vqe N4 & 28.66x & 639 & 188:186 & 1.01x \\
Vqe Uccsd N4 & 15.41x & 952 & 160:160 & 1.0x \\
Vqe Uccsd N8 & 4.48x & 33969 & 3816:3742 & 1.02x \\
Wstate N27 & 26.17x & 1704 & 416:416 & 1.0x \\
Wstate N3 & 71.54x & 67 & 14:14 & 1.0x \\
Wstate N76 & 3.42x & 2488 & 416:416 & 1.0x \\
\bottomrule
\end{tabular}
\caption{Clifford+T transpilation speedup using STABSim vs the ZX-Calculus optimizer in PyZX to achieve the same reduction~\cite{kissinger_reducing_2020}.  
\textbf{T:Opt} shows the initial and optimized number of T-type gates, and \textbf{T-reduction} shows the reduction factor.}
\label{tab:circuit-perf}
\end{table}

\end{document}